\documentclass[14pt]{article}
\usepackage{cite}
\usepackage{amssymb}
\usepackage{amsfonts}
\usepackage{mathrsfs}
\usepackage{graphicx}
\usepackage{amsmath}
\usepackage{color}
\usepackage{amsthm}
\usepackage{epstopdf}
\usepackage{booktabs}
\usepackage{pifont,bm}
\usepackage{tikz}
\usepackage{enumerate}
\usepackage{threeparttable}
\usepackage{epsfig}
\usepackage{subfigure}
\usepackage{float}
\usepackage{appendix}
\usepackage{relsize}
\usepackage{setspace}
\usepackage{multirow}

\theoremstyle{definition}

\makeatletter
\def\@biblabel#1{[#1]}
\makeatother

\makeatletter \@addtoreset{equation}{section}
\renewcommand{\theequation}{\arabic{section}.\arabic{equation}}

\begin{document}

\begin{titlepage}
\title{\bf { Integrable turbulence and statistical properties of chaotic wave fields
in the complex modified Korteweg--de Vries equation
\footnotetext{
\protect\\$\qquad^{\ast}$Corresponding author:
wqpeng@ouc.edu.cn (W.Q. Peng)}
}}

%
%
%

\author{Ya-Xin Luo, Xiao-Wang Lan, Wei-Qi Peng$^{\ast}$\\
\small \emph{School of Mathematical Sciences, Ocean University of China, Qingdao, 266100, China} \\
\date{}}
\thispagestyle{empty}
\end{titlepage}
\maketitle

\vspace{-0.5cm}
\begin{center}
\rule{15cm}{1pt}\vspace{0.3cm}

\parbox{15cm}{\normalsize
{\bf Abstract}\\
\hspace{0.5cm}
This paper investigates integrable turbulence and the mechanism for generating rogue waves in the complex modified Korteweg--de Vries (mKdV) equation through numerical simulations and statistical analysis. We solve the Lax pair eigenvalue problem using the Fourier collocation method and establish a correspondence among solitons, breathers, and rogue waves on the spectral plane. We generate chaotic wave fields by simulating the complex mKdV equation using the split-step Fourier method, starting from a plane wave with superimposed random noise. We analyze the effects of the standard deviation \(\mu\) and the correlation length \(L_c\). The results show that increasing \(\mu\) drives a transition from breather turbulence to soliton turbulence, increasing the probability of rogue-wave generation.In contrast, \(L_c\) primarily influences the spatial scale of the structures. As the constant background of the complex mKdV equation is modulationally stable, its integrable turbulence does not depend on modulational instability. Compared with the nonlinear Schr\"odinger (NLS) equation under identical initial conditions, the complex mKdV equation attains comparable amplitudes under strong perturbations, but exhibits a higher spatial gradient. This suggests that third-order dispersion and derivative nonlinearity alter the excitation mechanism and steepen the waveform.
}

\vspace{0.5cm}
\parbox{15cm}{\small{

\vspace{0.3cm} \emph{Key words:}  Integrable turbulence; Complex mKdV equation; Rogue wave; Chaotic wave field; Split-step Fourier method.\\

} }
\end{center}
\vspace{0.3cm} \rule{15cm}{1pt} \vspace{0.2cm}

\section{Introduction}\label{sec:intro}

Integrable turbulence refers to chaotic wave phenomena described by integrable nonlinear evolution equations \cite{Zakharov2009}. Unlike classical turbulence, integrable systems preserve an infinite number of conserved quantities during evolution, and energy can transfer among modes without dissipation. Depending on the dominant nonlinear structures in the wave field, integrable turbulence is typically classified into two regimes: soliton-dominated and breather-dominated turbulence \cite{Agafontsev2015, Akhmediev2016}. In recent years, with increasing attention to extreme events such as rogue waves,integrable turbulence---as a key background for rogue-wave generation---has become a research frontier in nonlinear science \cite{Dudley2014}.

The term "rogue wave" originated in oceanography to describe sudden, massive waves with peak amplitudes far exceeding those of the surrounding wave field. These waves have short lifetimes and are extremely localized. Researchers have since observed similar extreme wave events in optics, plasma physics, fluid dynamics, and ultracold atomic gases. In recent years, optical fiber experiments have enabled real-time, single-shot observation of rogue waves in integrable turbulence \cite{Suret2016, Walczak2015}, providing direct experimental support for related theoretical work. Extensive studies have shown that modulational instability (MI) is a key mechanism for the generation of rogue waves in NLS-type systems\cite{Onorato2006, Biondini2015, Dudley2014}, in which specific perturbation components in the initial noise are selectively amplified and grow exponentially, ultimately evolving into long-lived chaotic wave fields, i.e., integrable turbulence.

Integrable turbulence in NLS-type systems has been studied comparatively thoroughly. Zakharov pointed out that turbulence in integrable systems differs essentially from classical turbulence: the system preserves infinitely many conserved quantities, and energy is transferred reversibly among modes without dissipation \cite{Zakharov2009}. For the focusing NLS equation, integrable turbulence is usually excited by MI, whereby small perturbations on a plane-wave background are amplified exponentially and eventually develop into a chaotic wave field \cite{Zakharov2013}. Agafontsev and Zakharov then carried out a systematic numerical study of this process, obtaining the evolution of the probability density function of the wave intensity and revealing the formation and statistical properties of rogue waves \cite{Agafontsev2015}. On this basis, breather turbulence and soliton turbulence were further distinguished as two limiting regimes, and their difference was characterized through probability density functions and spectral features \cite{SotoCrespo2016,Akhmediev2016}. El
and Tovbis, in turn, provided a rigorous spectral description of soliton and breather gases \cite{El2020}. These studies established the intensity probability density function and the Lax eigenvalue spectrum as standard tools for characterizing integrable turbulence.

Building on this foundation, researchers have further extended their investigations to other integrable systems, such as the Hirota equation \cite{Wang2020Hirota}, coupled NLS systems \cite{Xie2020Coupled}, the fourth-order NLS equation \cite{Tang2022Fourth}, and the cubic--quintic NLS system \cite{Xie2026CubicQuintic}, among others. These studies have revealed how different dispersion and nonlinear structures affect turbulence characteristics and the probability of rogue wave generation. In recent years, research on integrable turbulence has also extended to the Kundu--Eckhaus (KE) equation \cite{Li2024KE} and the derivative nonlinear Schr\"{o}dinger equation \cite{Zhong2026}. The former examined the effects of the standard deviation and correlation length of initial perturbations on the statistical properties of the wave field; the latter, starting from the modulational instability of plane waves, revealed the asymmetric characteristics of their wave-action spectrum. However, these studies have primarily focused on NLS-type equations. For the complex mKdV equation---another important integrable model---integrable turbulence has not yet been systematically studied.

In this paper we focus on the complex modified Korteweg--de Vries (mKdV) equation,
\begin{equation}
q_t + q_{xxx} + 6|q|^2 q_x = 0,
\label{eq:mkdv}
\end{equation}
where \(q(x,t)\) is a complex-valued function. The second term of Eq.~\eqref{eq:mkdv} accounts for third-order dispersion, which reflects the difference in the propagation velocities of different frequency components, whereas the derivative nonlinearity describes the steepening of the waveform as it propagates. As a classical model in soliton theory, the complex mKdV equation is a completely integrable system possessing a Lax pair, infinitely many conservation laws and a rich family of exact solutions \cite{Ablowitz_Clarkson_1991}. Soliton, breather and higher-order rogue wave solutions can also be constructed by means of the Darboux transformation \cite{He2014cmKdV,He2013}. Since then, the exact solutions of the equation have been further generalized. By formulating the Riemann--Hilbert problem for the equation with vanishing boundary conditions, Zhang et al. obtained bound-state solitons from higher-order poles of the reflection coefficient and discussed their interaction with ordinary solitons \cite{Zhang2020Bound}. Liu et al. studied superregular breathers, showed that they are formed by the collision of quasi-Akhmediev breathers with a certain phase shift, and found that they can exhibit half-transition and full-suppression modes that do not exist in the standard NLS equation \cite{Liu2017}. Subsequently, the Darboux transformation was further employed to construct hybrid structures in which rogue waves and breathers coexist, with their interaction controllable through the parameters of the solution \cite{Ma2022}. A different line of research concerns dark solitons. Depending on the valley profile, the dark solitons of the complex mKdV equation fall into two types. Collisions between solitons of the same type preserve their shapes, whereas collisions between different types alter them \cite{Huang2026}. In contrast to these analytical studies, numerical work has focused on providing reliable discretizations of the equation, such as a Crank--Nicolson-type finite difference scheme with second-order accuracy and discrete mass conservation for the vanishing-boundary-value problem \cite{Xu2021}. These studies, however, have focused on the derivation of exact solutions and on the development of numerical methods, while the spectral features of the chaotic wave field, the transition of the turbulence type and the influence of the parameters remain to be explored in depth. It is worth emphasizing that, because of the derivative nonlinearity, a constant background of the complex mKdV equation is modulationally stable (see the derivation in Section~3.2), which makes the excitation and evolution mechanisms of its integrable turbulence fundamentally different from those of NLS-type equations.

This paper aims to fill this gap by systematically investigating the evolutionary characteristics of integrable turbulence in the complex mKdV equation. Accordingly, in Section~2, we use the Fourier collocation method to solve the Lax spectral problem and obtain spectral plots corresponding to different exact solutions. In Section~3, we perform numerical simulations using the split-step Fourier method to describe the process by which chaotic wave fields are generated from plane waves with superimposed random noise. We primarily analyze the effects of the standard deviation \(\mu\) and the correlation length \(L_c\) on the statistical properties of the wave field, including the probability density function and the wave-action spectrum, and compare and discuss the probability of rogue-wave generation with that of the NLS equation. Section~4 presents the conclusions and discussion.

\section{Spectral distributions of the exact solutions of the complex mKdV equation}
\label{sec:spectral}

As an integrable system, the complex mKdV equation possesses the Lax pair\cite{He2014cmKdV}
\begin{subequations}
  \begin{equation}
\psi_x = M\psi, \quad \psi_t = N\psi,
  \end{equation}
  \begin{equation}
  M = \begin{pmatrix}
-i\lambda & q \\
r & i\lambda
\end{pmatrix}, \quad
N = \sum_{j=0}^{3} V_j \lambda^j,\quad  V_{0}=\left(\begin{array}{cc}
     -qr_{x}+q_{x}r  &  -q_{xx}+2q^{2}r\\
     -r_{xx}+2qr^{2} &  qr_{x}-q_{x}r\\
 \end{array}\right),
  \end{equation}
  \begin{equation}
 V_{1}=\left(\begin{array}{cc}
     -2iqr  &  2iq_{x}\\
     -2ir_{x} &  2iqr\\
\end{array}\right),\quad V_{2}=\left(\begin{array}{cc}
     0  &  4q\\
     4r &  0\\
\end{array}\right),\quad V_{3}=\left(\begin{array}{cc}
    -4i  &  0\\
      0 &  4i\\
\end{array}\right),
  \end{equation}
\end{subequations}
where \(\psi=(\psi_{1},\psi_{2})^{T}\) is the vector eigenfunction, \(r = -q^{*}\) and \(\lambda\) is the spectral parameter. The compatibility condition of this Lax pair yields the complex mKdV equation.
To establish the correspondence between the exact solutions and the spectral distributions of the Lax pair, the bright soliton, the breather and the
rogue wave of the complex mKdV equation are presented below, together with their spatiotemporal evolutions and the corresponding eigenvalue
spectra.

\begin{figure}[H]
\centering
\noindent\includegraphics[width=4.8cm,height=4.6cm,angle=0]{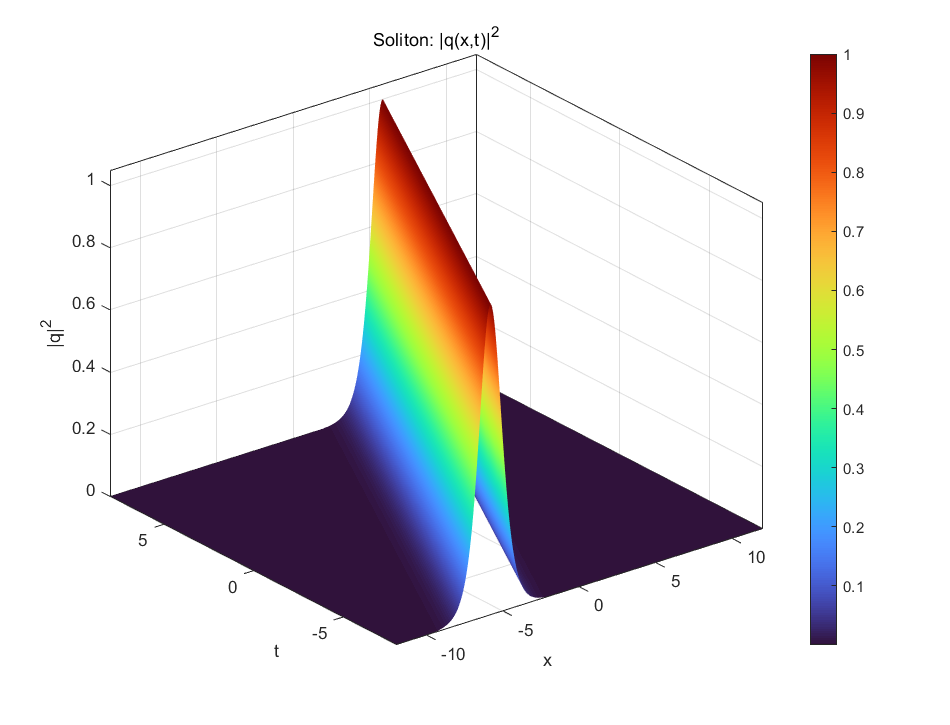}%
\hspace{0.8cm}%
\noindent\includegraphics[width=4.8cm,height=4.6cm,angle=0]{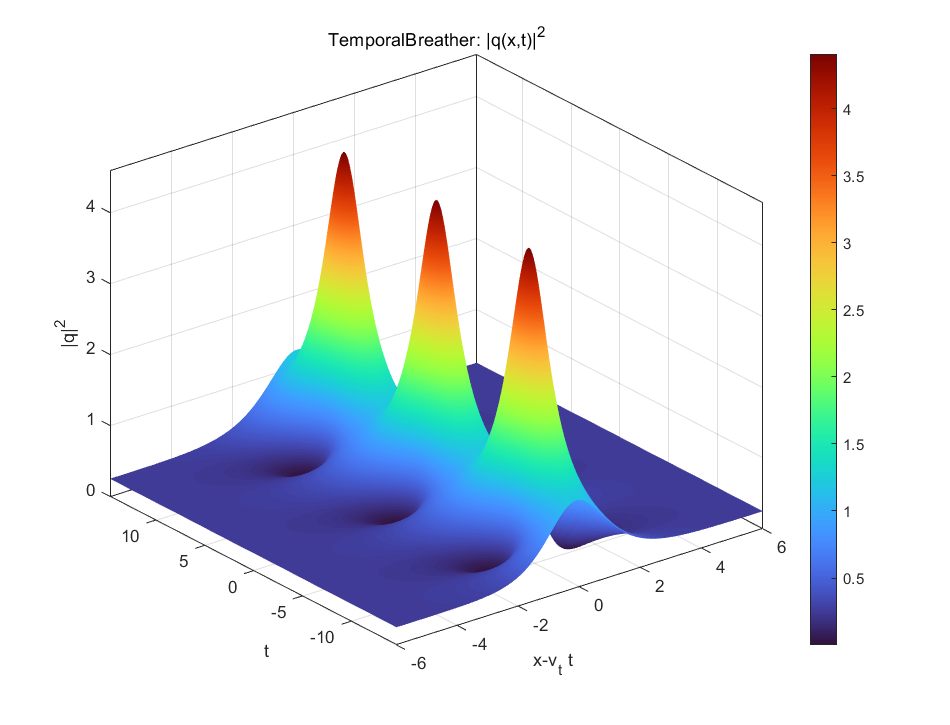}%
\par\smallskip
\noindent\parbox[t]{4.8cm}{\centering\small\strut\textbf{(a)} Bright soliton ($\xi=0.3$, $\eta=0.5$)\par}%
\hspace{0.8cm}%
\noindent\parbox[t]{4.8cm}{\centering\small\strut\textbf{(b)} Temporal breather ($c=0.5$, $a=0.3$, $\beta=0.8$)\par}%
\par
\medskip
\noindent\includegraphics[width=4.8cm,height=4.6cm,angle=0]{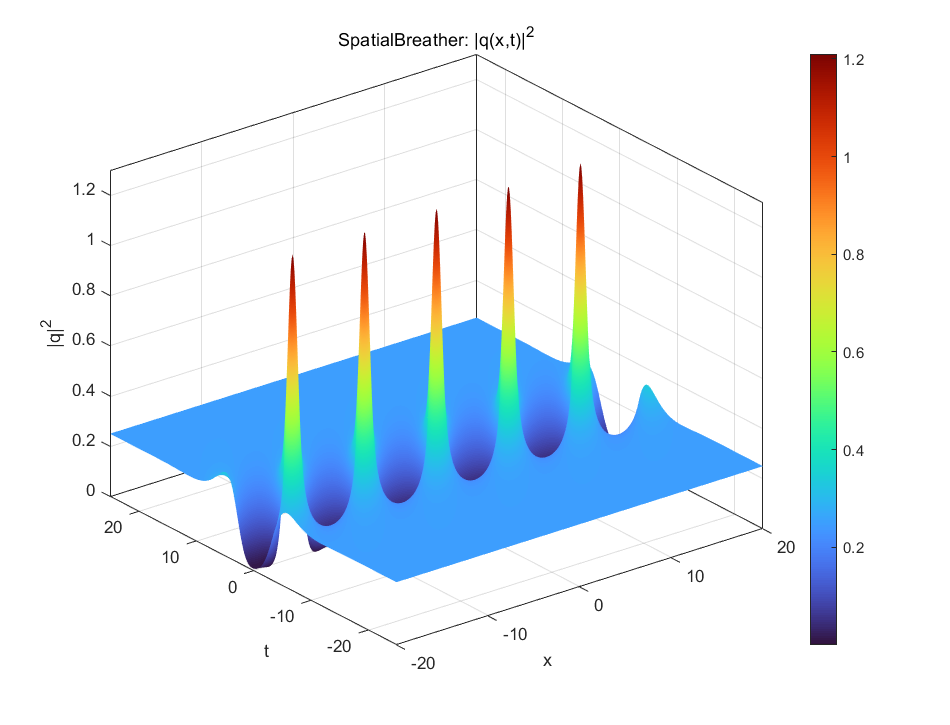}%
\hspace{0.8cm}%
\noindent\includegraphics[width=4.8cm,height=4.6cm,angle=0]{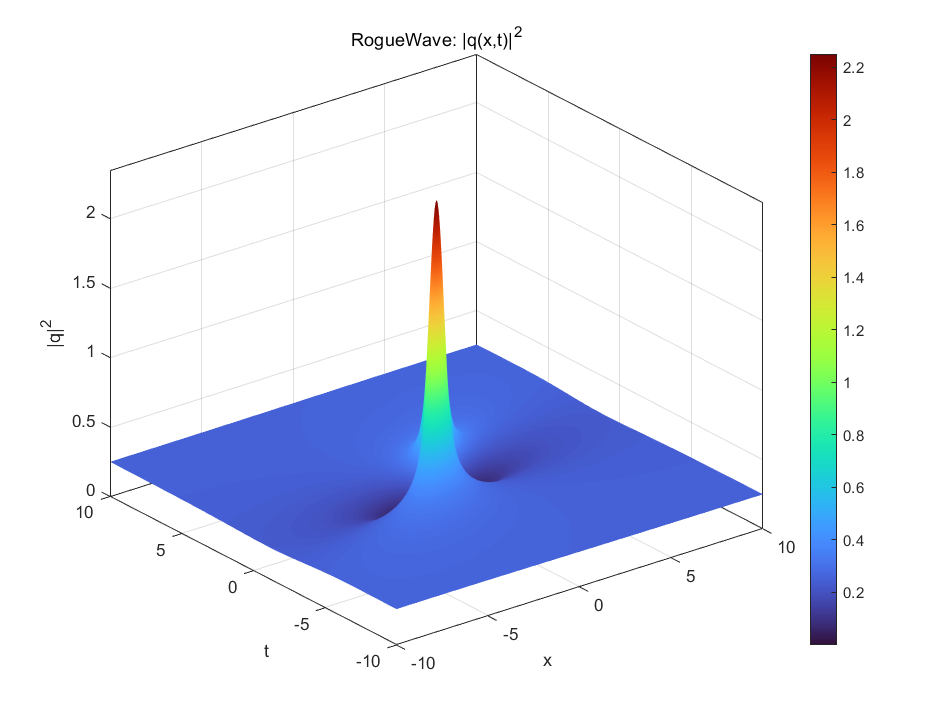}%
\par\smallskip
\noindent\parbox[t]{4.8cm}{\centering\small\strut\textbf{(c)} Spatial breather ($c=0.5$, $a=0.3$, $\beta=0.3$)\par}%
\hspace{0.8cm}%
\noindent\parbox[t]{4.8cm}{\centering\small\strut\textbf{(d)} Rogue wave ($c=0.5$, $a=0$)\par}%
\par
\caption{Three-dimensional spatiotemporal evolution of the wave-field intensity $|q|^{2}$ for the four analytical solutions of the complex mKdV equation.}
\label{fig:analytic_3d}
\end{figure}

\textbf{(1) Fundamental soliton solution}

The single-soliton solution of the complex mKdV equation can be written as \cite{Ablowitz_Clarkson_1991}
\begin{equation}
q(x,t) = 2\eta \,\mathrm{sech}\bigl[2\eta(x - (4\eta^2 - 3\xi^2)t-x_0)\bigr]\,
\cdot\exp\Bigl[i(\xi x -\xi(12\eta^2 - \xi^2)t + \theta_0)\Bigr],
\label{soliton}
\end{equation}
where the parameter $\eta$ determines the amplitude and the width of the soliton, and $\xi$ describes the carrier wavenumber of the soliton; together they
determine the propagation velocity $v=4\eta^{2}-3\xi^{2}$, while $x_0$ and $\theta_0$ denote the initial position and the initial phase, respectively.
Taking $\xi=0.3$ and $\eta=0.5$, the three-dimensional evolution of the soliton intensity is shown in Fig.~\ref{fig:analytic_3d}(a): the soliton keeps its
waveform unchanged during propagation, with a constant peak value $|q|^{2}=4\eta^{2}=1$, exhibiting the typical shape-preserving property.

From a spectral perspective, the soliton solution corresponds to a pair of complex-conjugate eigenvalues $\lambda = -\xi/2 \pm i\eta = -0.15 \pm 0.5i$ on the spectral plane of the Lax pair [Fig.~\ref{fig:analytic_spectra}(a2)]. These eigenvalues deviate from the imaginary axis. The imaginary part $\eta$ determines the soliton amplitude, while the real part $-\xi/2$ is associated with the carrier wavenumber and propagation velocity.

\textbf{(2) Single-breather solution}

Given the nonzero plane-wave seed solution $q_0(x,t)=ce^{i\rho}$, a single-breather solution of the complex mKdV equation can be constructed using the one-fold Darboux transformation \cite{He2014cmKdV}:
\begin{equation}
q(x,t)=c\,e^{i\rho}+4\beta\,\frac{\phi_1\bar{\phi}_2}{|\phi_1|^2+|\phi_2|^2},
\qquad
\phi_1=e^{i\rho/2}y_1,\quad \phi_2=e^{-i\rho/2}y_2,
\label{breather}
\end{equation}
where $c$ is the amplitude of the background plane wave, $a$ is the carrier wavenumber of the background, $\rho=ax+(a^3-6ac^2)t$, and $\beta$ is a real
parameter characterizing the breather, whose corresponding Darboux spectral parameter is $\lambda_{\mathrm{D}}=-a/2+i\beta$. The auxiliary quantities in
Eq.~\eqref{breather} are defined as
\begin{equation}
\begin{gathered}
\mu=\sqrt{c^2-\beta^2},\qquad
F=3a^2-2c^2-4\beta^2-6ia\beta,\\[4pt]
P=\mu\,(x+Ft),\\[4pt]
y_1=\cos P+\frac{\beta+c}{\mu}\sin P,\qquad
y_2=\cos P+\frac{\beta-c}{\mu}\sin P,
\end{gathered}
\label{breather_aux}
\end{equation}
where $\mu$ and $F$ are auxiliary constants determined by $a$, $c$ and $\beta$, $P$ is a complex phase variable, and $y_1$ and $y_2$ are auxiliary functions
determined by $P$ and are generally complex-valued. $\phi_1$ and $\phi_2$ are the corresponding eigenfunction components of the Darboux transformation.

The value of $\mu$ in Eq.~\eqref{breather_aux} determines the breather solution's classification. When $0<\beta<c$, $\mu$ is real; the solution is spatially periodic and temporally localized. This solution is called a spatially periodic breather (Akhmediev type). In contrast, when $\beta>c$, $\mu$ is purely imaginary, and the solution is spatially localized and temporally periodic. This case is called a temporally periodic breather (Kuznetsov--Ma type). As $\beta$ approaches $c$, $\mu$ approaches zero, and the solution reduces to a first-order rogue wave characterized by the critical Lax spectral point $\lambda=\pm ic$.

In this paper, we take $c=0.5$ and $a=0.3$, and set  $\beta=0.8$ and $\beta=0.3$, respectively. As shown in Fig.~\ref{fig:analytic_3d}(b), the temporally periodic breather appears as a spatially localized wave packet whose intensity varies periodically in time, first weakening and then strengthening. The spatially periodic breather in Fig.~\ref{fig:analytic_3d}(c) appears as a spatially periodic wave train whose intensity is markedly enhanced near the center, becomes localized in time, and approaches the background intensity $c^2=0.25$ as $t\to\pm\infty$.

In the gauge where the background carrier is removed, both breather types correspond to spectral points $\lambda=\pm i\beta$ on the imaginary axis of the
Lax spectral plane; the temporally periodic and the spatially periodic breather therefore correspond to $\lambda=\pm0.8i$ and $\lambda=\pm0.3i$, respectively [Figs.~\ref{fig:analytic_spectra}(b2) and \ref{fig:analytic_spectra}(c2)]. By contrast, the eigenvalues of the soliton solution are
$\lambda=-\xi/2\pm i\eta$; when $\xi\neq0$, their real part is nonzero and the eigenvalues therefore deviate from the imaginary axis.

\textbf{(3) Fundamental rogue wave solution}

Taking the limit $\beta\to c$ of the breather solution yields the first-order rogue wave of the complex mKdV equation \cite{He2014cmKdV,He2013}:
\begin{equation}
q^{[1]}(x,t)=-c\,e^{ia(x+t(a^{2}-6c^{2}))}\frac{A+48ic^{2}ta-3}{A+1},
\label{rogue}
\end{equation}
where
\begin{equation}
A=24ta^{2}c^{2}x+24ta^{2}c^{2}s_0+36t^{2}a^{4}c^{2}-48c^{4}tx-48c^{4}ts_0+8c^{2}xs_0+144c^{6}t^{2}+4c^{2}x^{2}+4c^{2}s_0^{2},
\end{equation}
with $a$, $c$ and $s_0$ being real parameters. This solution is doubly localized in space and time: as $x\to\infty$ or $t\to\infty$, $|q^{[1]}|^{2}\to c^{2}$, i.e., the height of the background is $c^{2}$, whereas a maximum peak value $9c^{2}$ is reached near the origin. Taking $c=0.5$, $a=0$ and
$s_0=0$, the three-dimensional evolution is shown in Fig.~\ref{fig:analytic_3d}(d), where the peak value reaches $|q|^{2}=2.25$, about nine times the background value $0.25$.

On the spectral plane, the rogue wave corresponds to the critical point $\lambda=\pm ic=\pm0.5i$ on the imaginary axis [Fig.~\ref{fig:analytic_spectra}(d2)]. This point is exactly the endpoint of the continuous spectrum of the background on the imaginary axis and also the limiting position of the breather eigenvalue $\pm i\beta$ as $\beta\to c$ \cite{He2013}. Therefore, the rogue wave is not a structure independent of the breather; rather, it is a critical configuration formed when the breather degenerates to the edge of the continuous spectrum on the spectral plane. This critical formation clarifies the generation mechanism of rogue waves from the spectral point of view.

For a given potential \(q(x,t)\), the spatial part of the Lax pair is recast into the linear eigenvalue problem
\begin{equation}
\begin{pmatrix}
-i\lambda & q \\
-q^* & i\lambda
\end{pmatrix}
\begin{pmatrix}
\psi_1 \\ \psi_2
\end{pmatrix}
=
\begin{pmatrix}
\psi_{1,x} \\ \psi_{2,x}
\end{pmatrix}.
\label{eq:eig}
\end{equation}

The spectral distributions corresponding to the various exact solutions are obtained by exact solutions, Eq.~\eqref{eq:eig} is solved by means of the Fourier collocation method \cite{Yang2010}. To this end the \(x\) axis is truncated to the finite interval \([-L/2,L/2]\), and \(\psi_1\), \(\psi_2\), and $q$ are expanded in terms of Fourier series:

\begin{figure}[p]
\centering
\noindent\includegraphics[width=4.8cm,height=4.0cm,angle=0]{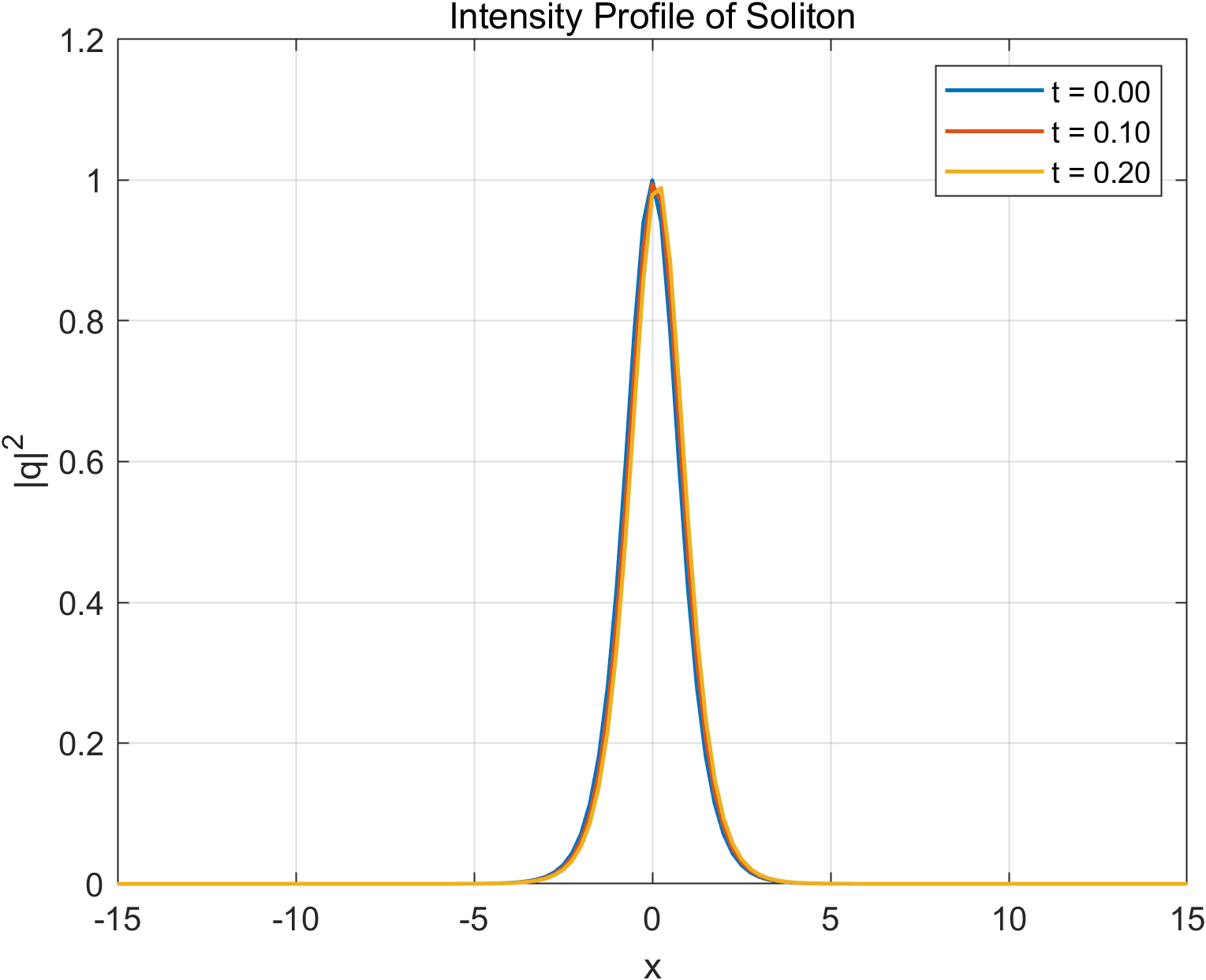}%
\hspace{0.8cm}%
\noindent\includegraphics[width=4.8cm,height=4.0cm,angle=0]{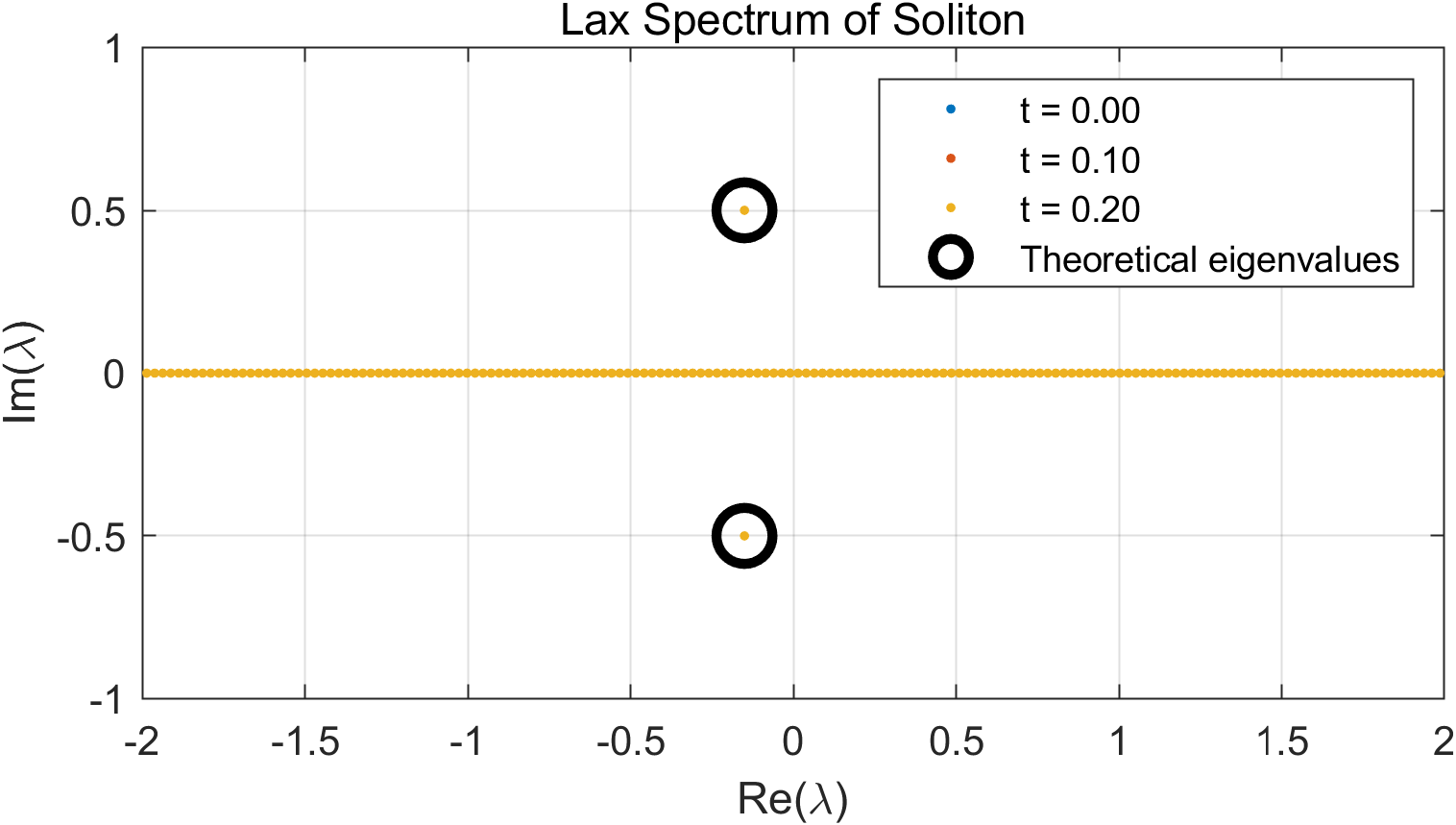}%
\par\smallskip
\noindent\parbox[t]{4.8cm}{\centering\small\strut\textbf{(a1)} Bright soliton\par}%
\hspace{0.8cm}%
\noindent\parbox[t]{4.8cm}{\centering\small\strut\textbf{(a2)} Bright soliton ($\lambda=-0.15\pm0.5i$)\par}%
\par
\medskip
\noindent\includegraphics[width=4.8cm,height=4.0cm,angle=0]{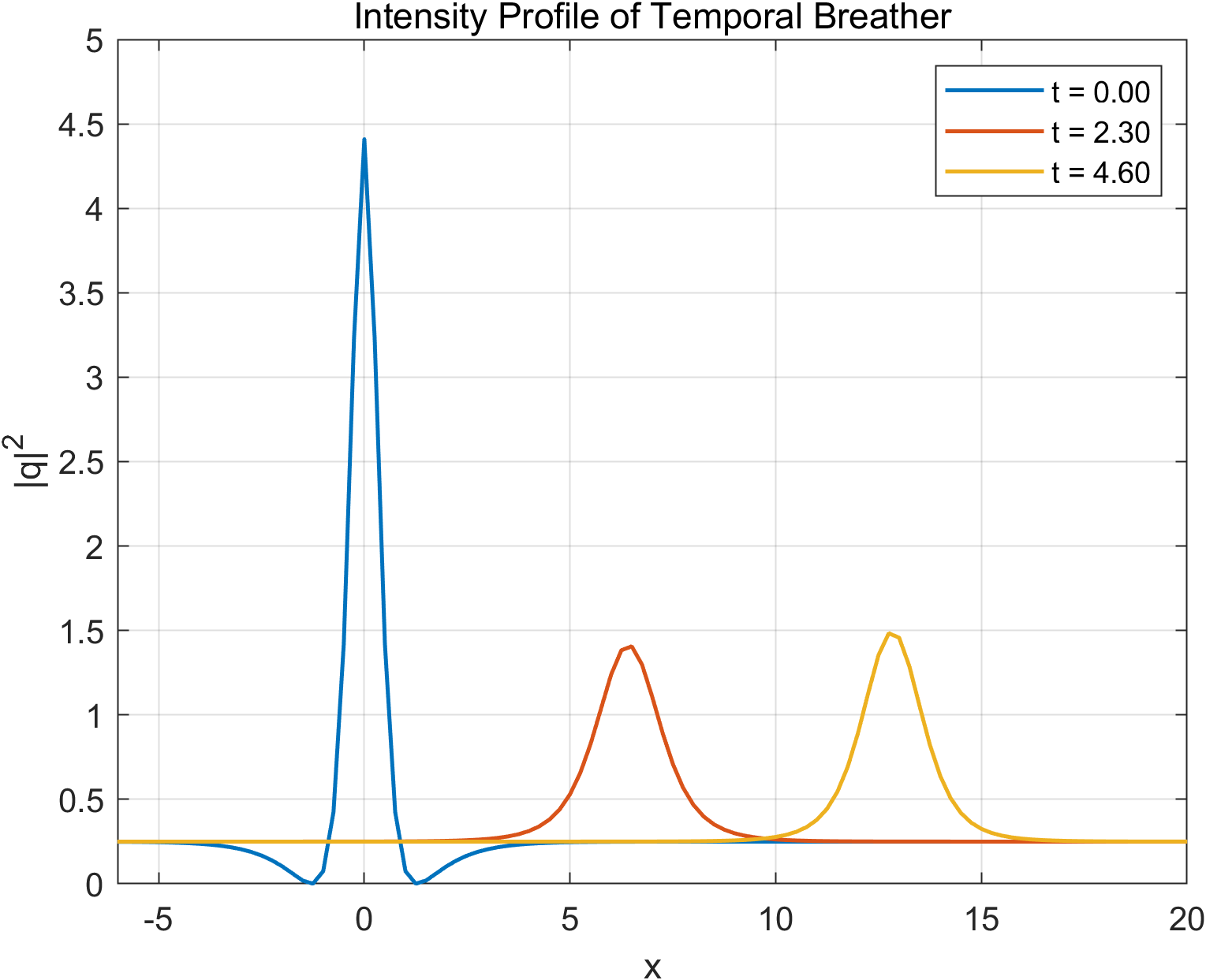}%
\hspace{0.8cm}%
\noindent\includegraphics[width=4.8cm,height=4.0cm,angle=0]{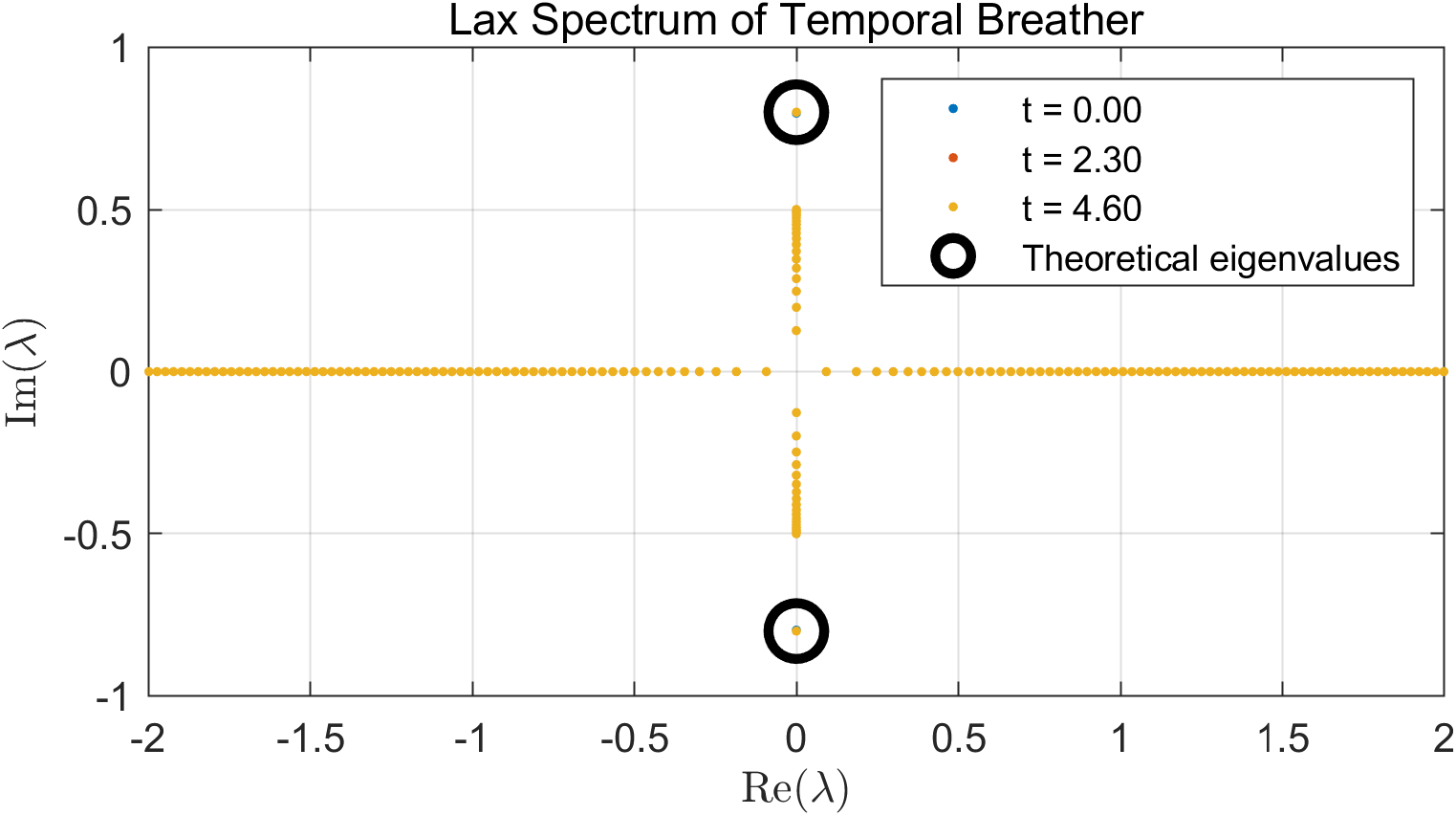}%
\par\smallskip
\noindent\parbox[t]{4.8cm}{\centering\small\strut\textbf{(b1)} Temporal breather\par}%
\hspace{0.8cm}%
\noindent\parbox[t]{4.8cm}{\centering\small\strut\textbf{(b2)} Temporal breather ($\lambda=\pm0.8i$)\par}%
\par
\medskip
\noindent\includegraphics[width=4.8cm,height=4.0cm,angle=0]{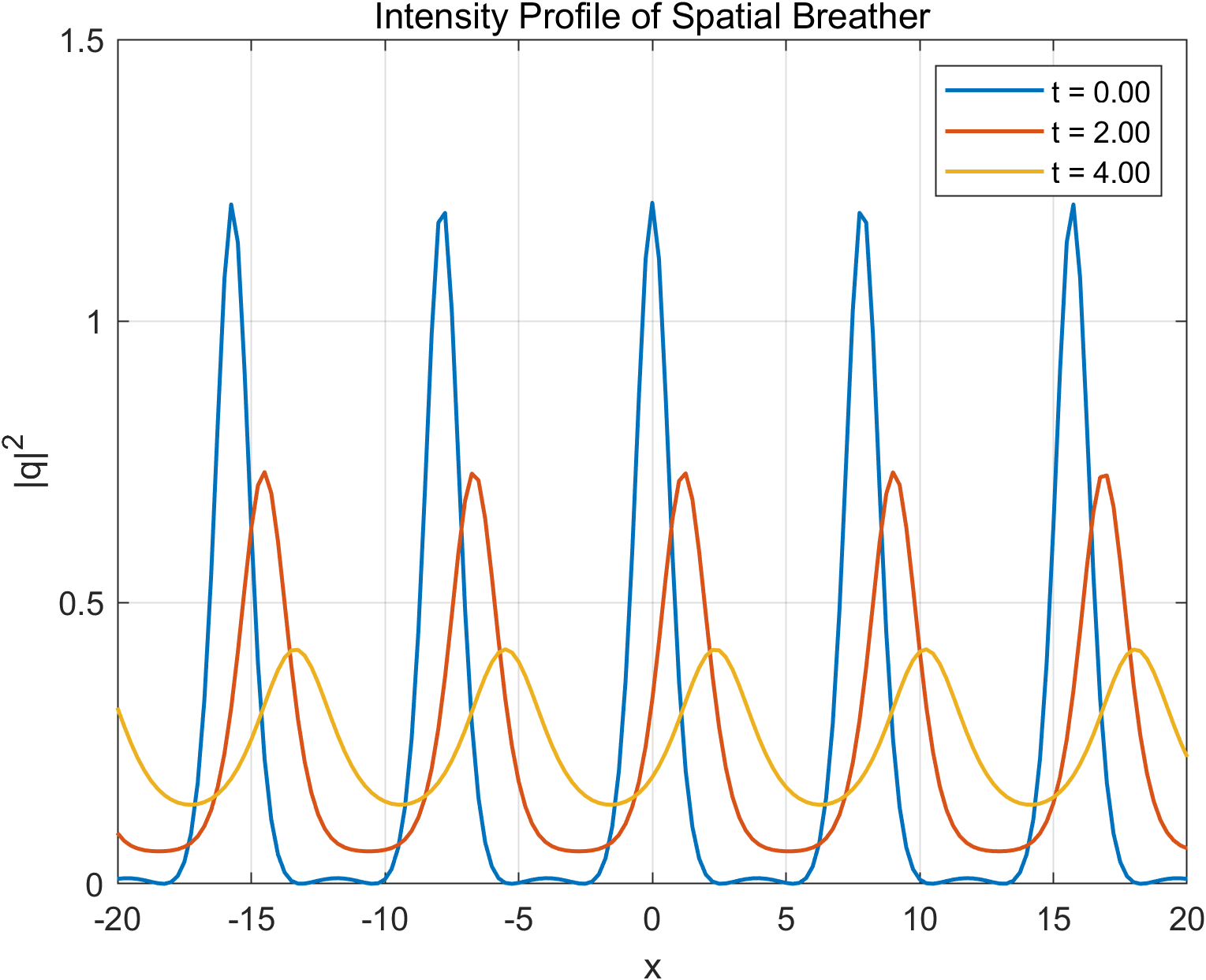}%
\hspace{0.8cm}%
\noindent\includegraphics[width=4.8cm,height=4.0cm,angle=0]{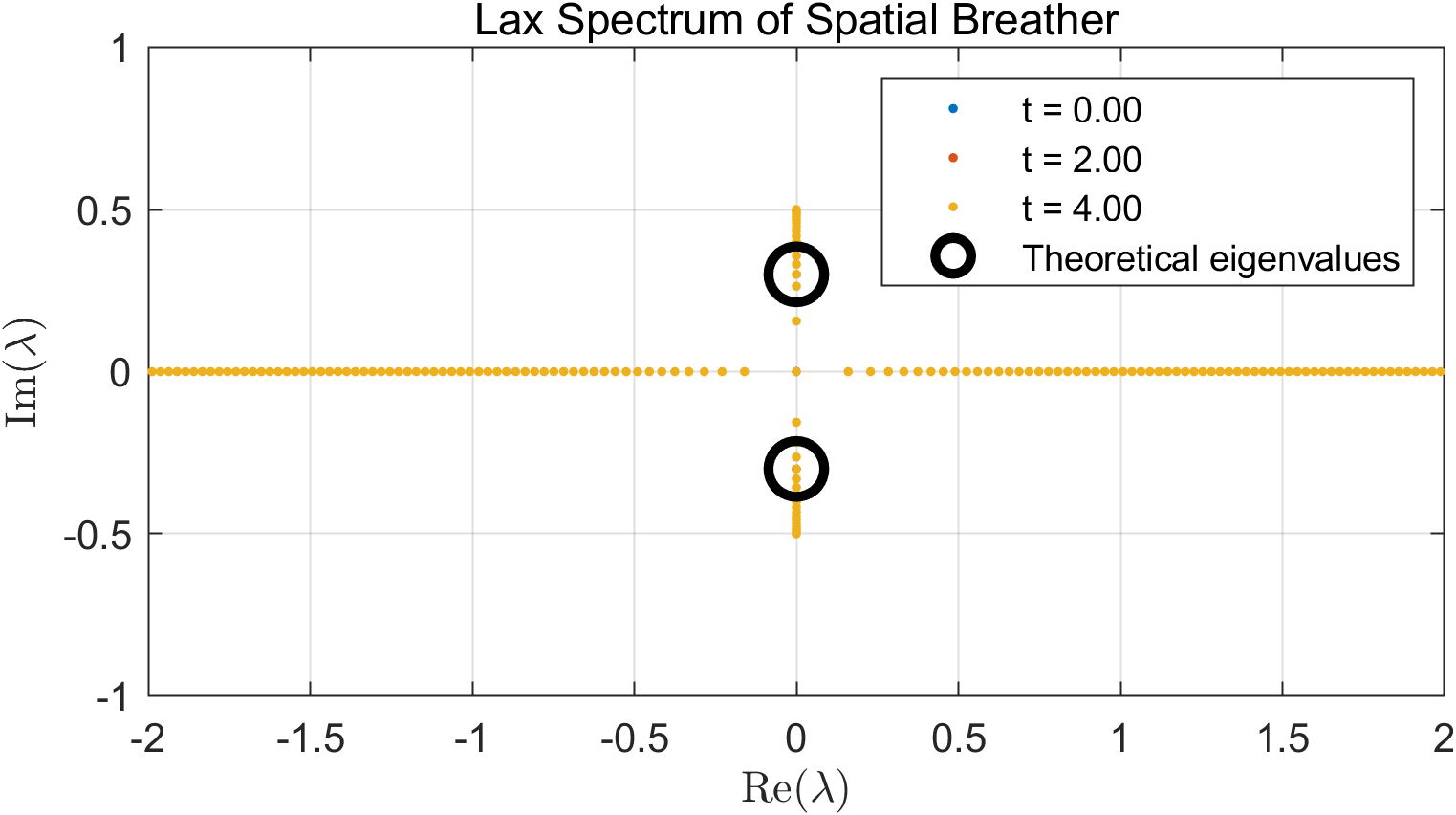}%
\par\smallskip
\noindent\parbox[t]{4.8cm}{\centering\small\strut\textbf{(c1)} Spatial breather\par}%
\hspace{0.8cm}%
\noindent\parbox[t]{4.8cm}{\centering\small\strut\textbf{(c2)} Spatial breather ($\lambda=\pm0.3i$)\par}%
\par
\medskip
\noindent\includegraphics[width=4.8cm,height=4.0cm,angle=0]{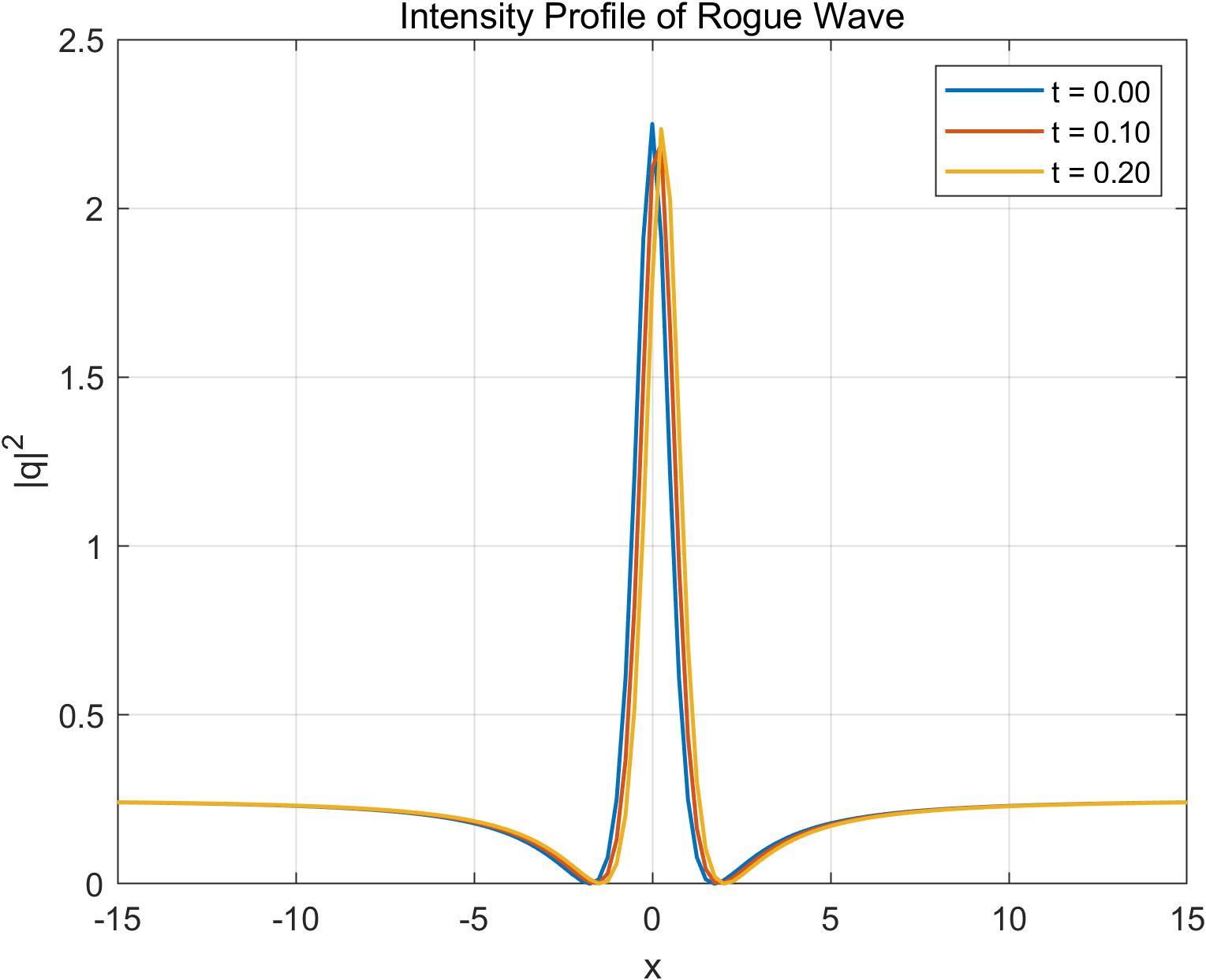}%
\hspace{0.8cm}%
\noindent\includegraphics[width=4.8cm,height=4.0cm,angle=0]{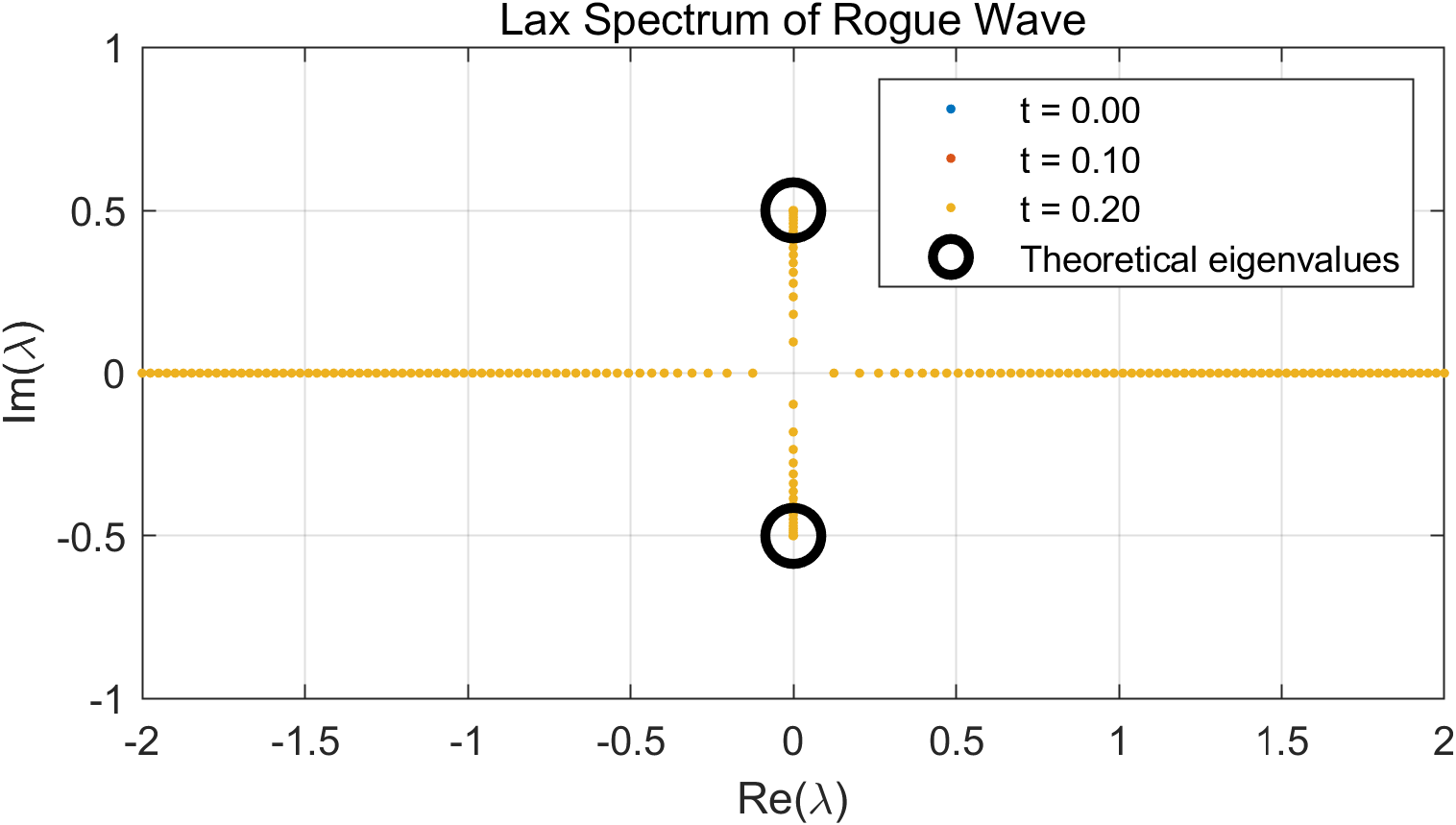}%
\par\smallskip
\noindent\parbox[t]{4.8cm}{\centering\small\strut\textbf{(d1)} Rogue wave\par}%
\hspace{0.8cm}%
\noindent\parbox[t]{4.8cm}{\centering\small\strut\textbf{(d2)} Rogue wave ($\lambda=\pm0.5i$)\par}%
\par
\caption{Intensity profiles of the four analytical solutions (left column, evaluated at three times) and the corresponding eigenvalue spectra of the Lax pair (right column). Black circles denote the theoretical eigenvalues.}
\label{fig:analytic_spectra}
\end{figure}
\clearpage

\begin{equation}
\psi_1 = \sum_{n=-N}^{N} a_{1n} e^{i n k_0 x},\quad
\psi_2 = \sum_{n=-N}^{N} a_{2n} e^{i n k_0 x},\quad
q = \sum_{n=-N}^{N} b_n e^{i n k_0 x},\quad k_0 = \frac{2\pi}{L}.
\end{equation}
Substituting these expansions into Eq.~\eqref{eq:eig} and collecting coefficients for the same Fourier modes converts the problem into a matrix eigenvalue problem, with eigenvalues that provide a numerical approximation of \(\lambda\).

The left column of Fig.~\ref{fig:analytic_spectra} shows the intensity profiles $|q|^{2}$ of the four exact solutions at three different times. The bright soliton [Fig.~\ref{fig:analytic_spectra}(a1)] is a single smooth isolated peak, and the profiles at the three times coincide exactly, which further confirms the shape-preserving property. The temporally periodic breather [Fig.~\ref{fig:analytic_spectra}(b1)] is a spatially localized wave packet whose peak moves along the positive $x$ direction; its height first decreases from about $4.4$ to about $1.4$ and then increases again. The spatially periodic breather [Fig.~\ref{fig:analytic_spectra}(c1)] is a spatially equidistant wave train with a peak separation of about $8$, in which the peaks gradually decrease, and the troughs are continuously raised, so the oscillation becomes progressively flatter. The rogue wave [Fig.~\ref{fig:analytic_spectra}(d1)] appears as a single sharp peak on the background, with zero-intensity troughs on each side of the peak.

The right column shows the corresponding eigenvalue spectra of the Lax pair, in which the black circles mark the theoretical eigenvalues of the respective exact solutions. Comparing these spectra reveals that the four solutions have clearly distinct counterparts on the spectral plane: the soliton corresponds to a pair of complex-conjugate eigenvalues with nonzero real part, whereas the rogue wave corresponds to the critical point $\lambda=\pm ic$ on the imaginary axis, which is also the endpoint of the continuous spectral segment on the imaginary axis. The eigenvalues of both breather types also lie on the imaginary axis, but their positions relative to the continuous spectrum differ. A constant background possesses a continuous spectral segment $[-ic, ic]$ on the imaginary axis. When $\beta>c$ (temporally periodic breather), the eigenvalue $\pm i\beta$ lies outside this segment and is an isolated discrete spectral point; when $0<\beta<c$ (spatially periodic breather), $\pm i\beta$ lies inside this segment and is embedded in the continuous spectrum. These spectral features provide a theoretical basis for identifying soliton and breather components in a chaotic wave field.

Moreover, the eigenvalues at the three different times in Figs.~\ref{fig:analytic_spectra}(a2)--(d2) essentially coincide. According to inverse scattering theory, the spectral parameter \(\lambda\) remains invariant during the evolution \cite{Lax1968, ZakharovShabat1972, Ablowitz_Clarkson_1991}. Consequently, the type and the number of nonlinear structures in a wave field can be inferred from the spectral distribution at the initial time \cite{Li2024KE, SotoCrespo2016, Akhmediev2016}. We use this property in Section~3 to analyze the composition of chaotic wave fields under different parameter conditions.

\section{Numerical simulation and statistical analysis of chaotic wave fields in the complex mKdV equation}\label{sec:numerical}
\subsection{Numerical method}
\label{sec:num_method}

The complex mKdV equation is solved numerically using the split-step Fourier (SSF) method \cite{Yang2010}, which decomposes the equation into linear and nonlinear parts. The linear part of Eq.~\eqref{eq:mkdv} is given by
\begin{equation}
q_t = -q_{xxx},
\label{eq:linear}
\end{equation}
which can be solved exactly in Fourier space, whereas the nonlinear part satisfies
\begin{equation}
q_t = -6|q|^2 q_x.
\label{eq:nonlinear}
\end{equation}

Let the spatial interval be $[-L/2, L/2]$ with grid points $X_j=-L/2+j\Delta x$ ($j=0,1,\dots,N-1$) and $\Delta x=L/N$. Denoting by $Q_j(t)$ the approximation of $q(X_j,t)$, each time step consists of two substeps. The linear part is treated first: a discrete Fourier transform is applied to the current $Q_j^m$, each transformed component is multiplied by $\exp(ik^3\Delta t)$, and an inverse Fourier transform is then performed to return to the original grid points. Thus, the update is given by
\begin{equation}
Q_j^{m+1}=\mathcal{F}^{-1}[\mathcal{F}(Q_j^m)\exp(ik^3\Delta t)],
\label{eq:ssf_linear}
\end{equation}
where $\mathcal{F}$ and $\mathcal{F}^{-1}$ denote the discrete forward and inverse Fourier transforms, $\Delta t$ is the time step, and $k$ is the wavenumber. For the nonlinear part~\eqref{eq:nonlinear}, the spatial derivative $q_x$ is computed by means of the Fourier transform, which yields an ordinary differential equation at each grid point,
\begin{equation}
\dot{Q_j}=-6|Q_j|^2\mathcal{F}^{-1}[ik\mathcal{F}(Q_j)].
\label{eq:ssf_nonlinear}
\end{equation}
Here $\dot{Q_j}$ denotes the derivative with respect to time $t$. In this context, Equation~\eqref{eq:ssf_nonlinear} is integrated in time by means of the fourth-order Runge--Kutta method, which keeps the conserved quantities essentially unchanged \cite{Trefethen2000}.

In the present simulations, we use $N=912$ spatial grid points, a spatial interval of length $L=100$, a time step $\Delta t=10^{-5}$, and a total evolution time $T=100$. To verify convergence with respect to the time step, we repeat the computation with $\Delta t$ halved to $5\times10^{-6}$. The evolution curves of the maximum amplitude and the probability density function essentially coincide, showing that the results have converged.

A different set of discretization parameters ($N=864$, $L=200$) is used for the computation of the eigenvalue spectra, so as to obtain a finer wavenumber resolution over a longer spatial interval.

All statistical quantities in this paper, including the maximum amplitude, the probability density function, the eigenvalue spectra, and the wave-action spectrum, are computed from a single random realization. Since the computational domain is sufficiently large ($L=100$), this single simulation is sufficient to capture the statistical characteristics of the wave field.

\subsection{Evolution of the chaotic wave field and influence of the parameters}
\label{sec:evolution}

To excite integrable turbulence, a plane wave with superimposed random noise is adopted as the initial condition,
\begin{equation}
q(x,0) = c + \mu f(x),
\label{initial}
\end{equation}
where $c=1$ represents the amplitude of the background plane wave, $\mu$ is the overall standard deviation of $q(x,0)$, and $f(x)$ is a complex-valued random function generated by convolving complex Gaussian white noise with a Gaussian kernel, where the kernel function is given by
\begin{equation}
h(x_k)=\sqrt{\frac{2\Delta x}{\sqrt{\pi}\,L_c}}\,
\exp\!\left[-2\left(\frac{x_k}{L_c}\right)^2\right],
\label{eq:kernel}
\end{equation}
with $x_k=k\Delta x$ denoting the relative offset, the coefficient in Eq.~\eqref{eq:kernel} ensures that the standard deviation of $f(x)$ after the convolution is $\sigma=1$, while the kernel width is determined by the correlation length $L_c$. The larger $L_c$ is, the more gently $f(x)$ varies in space and the larger the separation between adjacent peaks becomes \cite{Akhmediev2009, SotoCrespo2016}. Under this initial condition, a finite-amplitude random perturbation is sufficient to trigger the chaotic evolution of the system; in NLS-type systems this process is governed by modulational instability \cite{SotoCrespo2016, Akhmediev2016}. Since the initial field is controlled only by $\mu$ and \(L_c\), the influence of these two parameters on the integrable turbulence of Eq.~\eqref{eq:mkdv} is analyzed below.

It should be emphasized, however, that the amplification of the wave field observed here is not caused by modulational instability.

The evidence comes from a linear stability analysis. Let the constant background be $q_0=c$, and superimpose a small perturbation $q=c+\epsilon v$. Retain only terms linear in $\epsilon$ after substituting into Eq.~\eqref{eq:mkdv}. The nonlinear term $6|q|^2q_x$ then splits into three parts: one part is $6|q_0|^2v_x=6c^2v_x$, and the other two are $6q_0q_{0x}\bar v$ and $6\bar q_0q_{0x} v$; since a constant background satisfies $q_{0x}=0$, the latter two parts vanish identically, and the perturbation equation reduces to
\begin{equation}
v_t+v_{xxx}+6c^2v_x=0.
\end{equation}
This is a linear equation with constant coefficients, so its elementary solutions can be obtained directly. Substituting a sinusoidal wave of the form $v(x,t)=A\,e^{i(kx-\omega t)}$ and canceling the common factor gives the relation that the frequency $\omega$ and the wavenumber $k$ must satisfy, as follows:
\begin{equation}
\omega=6c^2k-k^3 .
\end{equation}
Whether $\omega$ is real or complex is decisive for the stability of the perturbation: if $\omega$ is real, $\exp(-i\omega t)$ is merely a phase factor and the perturbation only propagates with an unchanged amplitude; only when $\omega$ acquires an imaginary part does the perturbation grow or decay exponentially in time, which corresponds to an unstable system. Since the right-hand side of the above expression is real for any real $k$, perturbations of all wavelengths can only propagate and never grow. Thus, for real $k$, a constant background is modulationally stable.

This can also be verified directly by numerical simulation: for $\mu=10^{-3}$, the maximum amplitude of the wave field remains at 1.002 within $t\in[0,20]$ without any growth. This phenomenon is in sharp contrast to the NLS \cite{SotoCrespo2016, Akhmediev2016} and KE \cite{Li2024KE} equations, whose constant backgrounds are modulationally unstable.

Therefore, the amplitude growth observed in this work does not originate from linear instability. Instead, because the initial perturbation amplitude is not small, with $\mu$ reaching a maximum of 0.6, the nonlinear terms take effect directly from the outset, causing the waveform to become locally steeper and energy to concentrate at a few locations, ultimately forming a spike structure with an amplitude far exceeding the background.

\begin{figure}[H]
\centering
\includegraphics[width=7.2cm,height=5.2cm,angle=0]{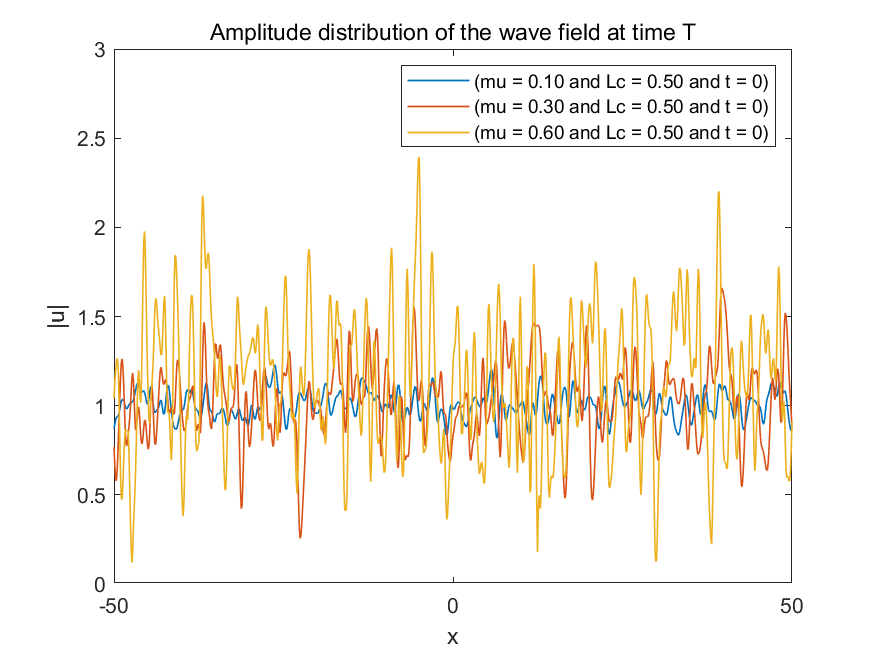}%
\hspace{0.3cm}%
\includegraphics[width=7.2cm,height=5.2cm,angle=0]{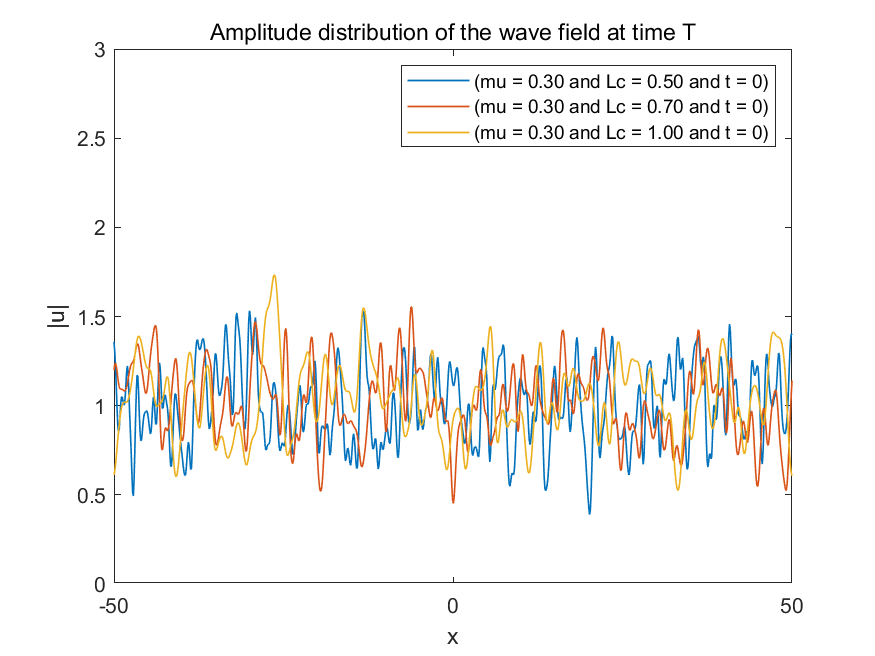}%
\par\smallskip
\parbox[t]{7.2cm}{\centering\small\strut\textbf{(a)} Initial field distribution for different $\mu$ ($L_c=0.5$)\par}%
\hspace{0.3cm}%
\parbox[t]{7.2cm}{\centering\small\strut\textbf{(b)} Initial field distribution for different $L_c$ ($\mu=0.30$)\par}%
\par
\caption{Spatial distribution of the initial wave-field intensity $|q(x,0)|^{2}$ for different parameters.}
\label{fig:initial}
\end{figure}

Figure~\ref{fig:initial}(a) shows the spatial distribution of the initial field intensity at a fixed $L_c=0.5$ for different values of $\mu$. For $\mu=0.10$, the field intensity fluctuates only slightly around the unit background value, indicating a weak initial perturbation and a slow development of the nonlinear amplification, so that the wave field requires a long time to enter the chaotic state. When $\mu$ increases to $0.30$ and even $0.60$, the deviation of the wave field from the background is markedly intensified, and the random fluctuations are significantly enhanced, so that the wave field enters the turbulent state more rapidly and reaches a higher peak amplitude in the statistically stationary state. Figure~\ref{fig:initial}(b) shows the difference in the initial field distribution at a fixed $\mu=0.30$ for different $L_c$. An increase in $L_c$ markedly lowers the spatial oscillation frequency of the fluctuations, which manifests itself as a gradual increase in the separation between adjacent peaks, but the extreme amplitudes of the curves do not change appreciably with $L_c$. This indicates that the correlation length mainly affects the spatial density of the noise rather than the magnitude of its deviation from the background field.

\begin{figure}[H]
\centering
\includegraphics[width=7.2cm,height=5.2cm,angle=0]{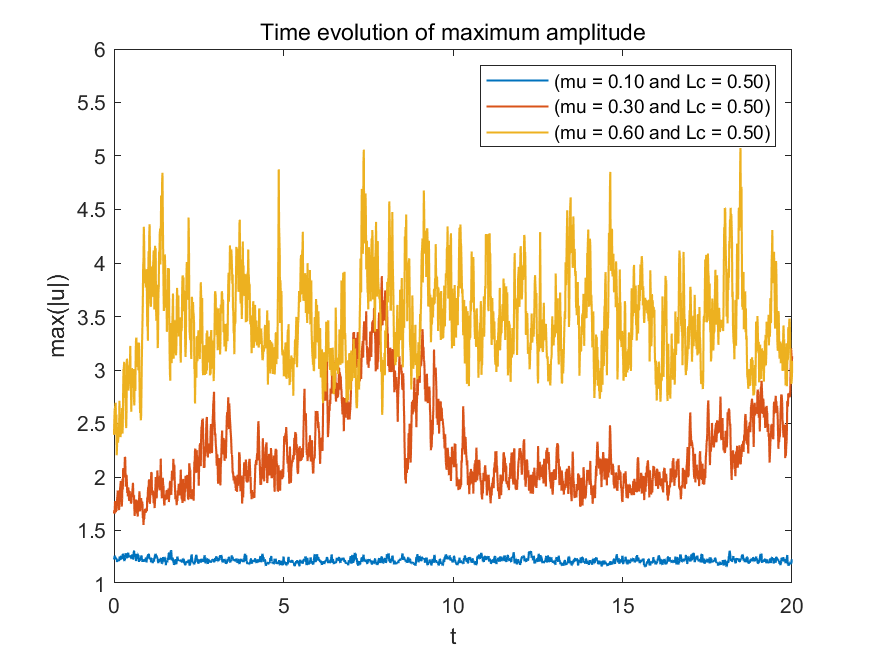}%
\hspace{0.3cm}%
\includegraphics[width=7.2cm,height=5.2cm,angle=0]{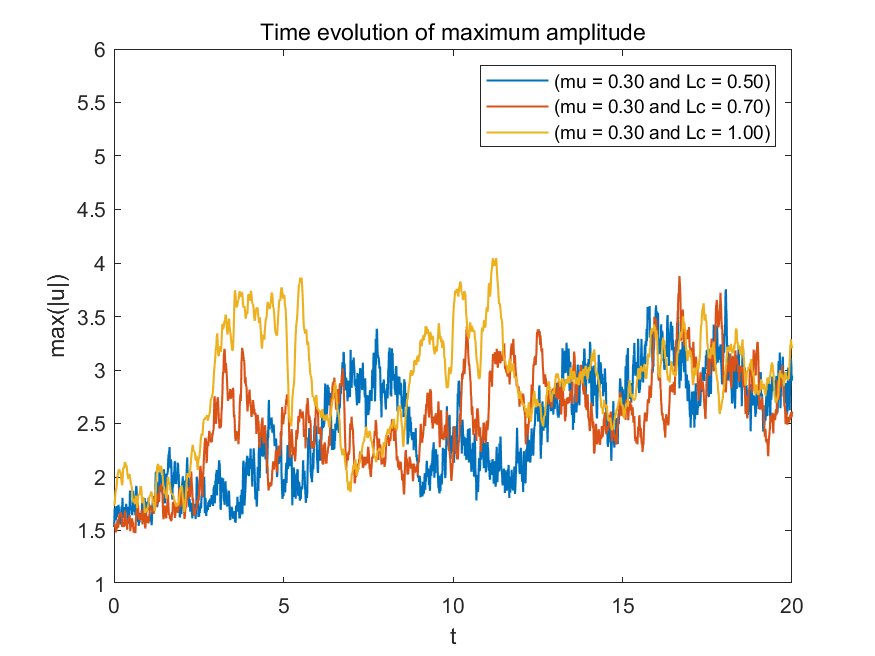}%
\par\smallskip
\parbox[t]{7.2cm}{\centering\small\strut\textbf{(a)} Maximum-amplitude evolution for different $\mu$ ($L_c=0.5$)\par}%
\hspace{0.3cm}%
\parbox[t]{7.2cm}{\centering\small\strut\textbf{(b)} Maximum-amplitude evolution for different $L_c$ ($\mu=0.30$)\par}%
\par
\caption{Temporal evolution of the maximum wave-field amplitude $|q|_{\max}$ for different parameters.}
\label{fig:maxamp}
\end{figure}

The evolution of the maximum amplitude $|q|_{\max}$ with time is examined next for different values of $\mu$, as shown in Fig.~\ref{fig:maxamp}(a). For each value of $\mu$, the maximum amplitude evolves significantly. When $\mu=0.1$, it fluctuates slightly around its initial level (about 1.2) without any appreciable growth. When $\mu$ increases to 0.3 and 0.6, the maximum amplitude rises rapidly in the initial stage (from about 1.6 and 2.4 to about 2.5 and 3.5, respectively), after which the growth slows down, and a plateau of oscillations is reached; eventually, the amplitude fluctuates irregularly around a certain mean value, indicating that a chaotic wave field has formed.

Furthermore, as the setup changes with increasing $\mu$, the wave field enters the chaotic state more quickly, and the mean amplitude level of that state also rises: for $\mu=0.1$ the amplitude remains at about $1.2$, whereas for $\mu=0.6$ it rises to about $3.5$ with peaks up to about $5$. This shows that the strength of the initial perturbation determines both the rapidity with which the wave field enters the chaotic state and the level of the amplitude in that state.

Figure~\ref{fig:maxamp}(b) shows the evolution of the maximum amplitude for different $L_c$ at a fixed $\mu=0.30$. The three curves all start from about 1.5 and grow gradually, and after $t\approx15$ they converge to a similar level of about $2.5$--$3$. The curve for $L_c=1.00$ rises fastest in the early stage and has the highest peak, i.e., for larger $L_c$ the wave field converges more rapidly towards the chaotic state and the amplitude fluctuations in the transitional stage are more pronounced. Compared with Fig.~\ref{fig:maxamp}(a), however, the three curves for different $L_c$ approach each other closely at later times with little difference, whereas the curves for different $\mu$ remain separated throughout with pronounced differences in level. The influence of $L_c$ on the evolution of the wave field is therefore more moderate than that of $\mu$.

\begin{figure}[H]
\centering
\noindent\includegraphics[width=0.32\linewidth,height=4.6cm,angle=0]{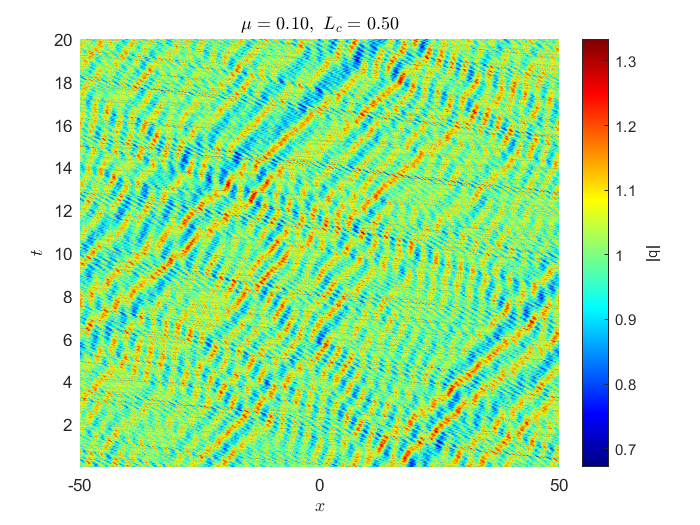}%
\hfill%
\noindent\includegraphics[width=0.32\linewidth,height=4.6cm,angle=0]{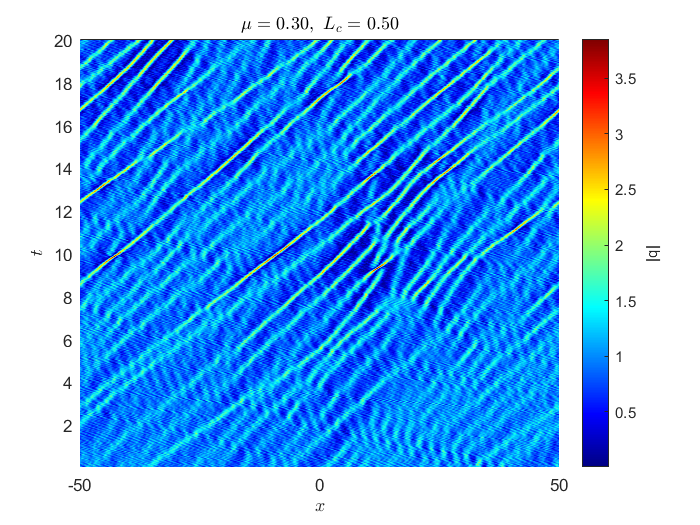}%
\hfill%
\noindent\includegraphics[width=0.32\linewidth,height=4.6cm,angle=0]{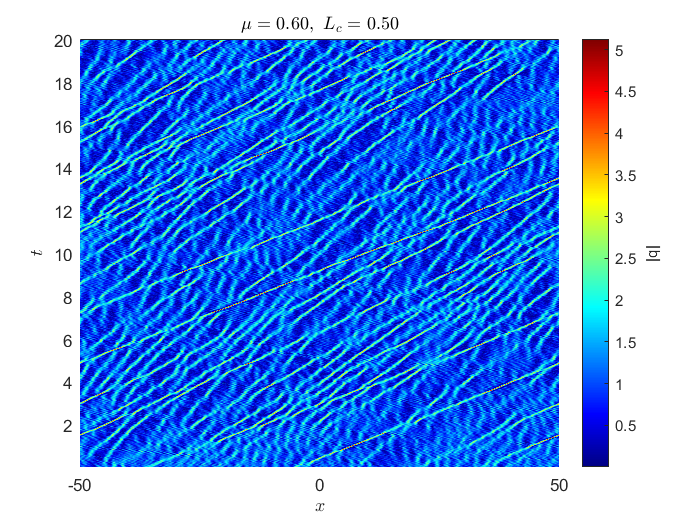}%
\par\smallskip
\noindent\parbox[t]{0.32\linewidth}{\centering\small\strut\textbf{(a)} $\mu=0.10$, $L_c=0.50$\par}%
\hfill%
\noindent\parbox[t]{0.32\linewidth}{\centering\small\strut\textbf{(b)} $\mu=0.30$, $L_c=0.50$\par}%
\hfill%
\noindent\parbox[t]{0.32\linewidth}{\centering\small\strut\textbf{(c)} $\mu=0.60$, $L_c=0.50$\par}%
\par
\medskip
\noindent\includegraphics[width=0.32\linewidth,height=4.6cm,angle=0]{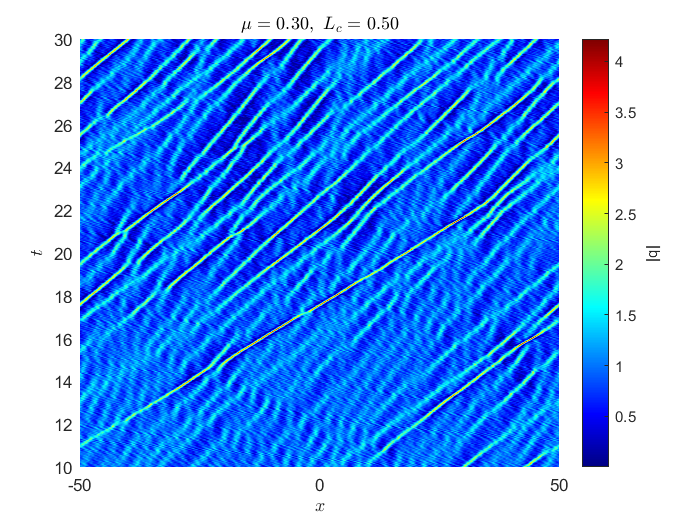}%
\hfill%
\noindent\includegraphics[width=0.32\linewidth,height=4.6cm,angle=0]{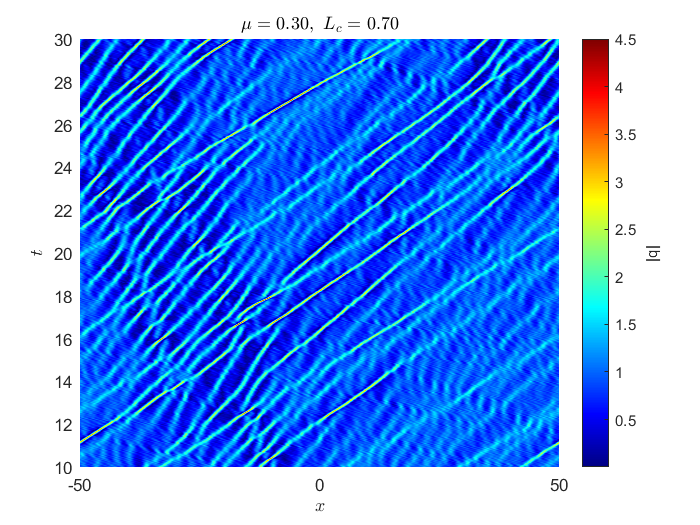}%
\hfill%
\noindent\includegraphics[width=0.32\linewidth,height=4.6cm,angle=0]{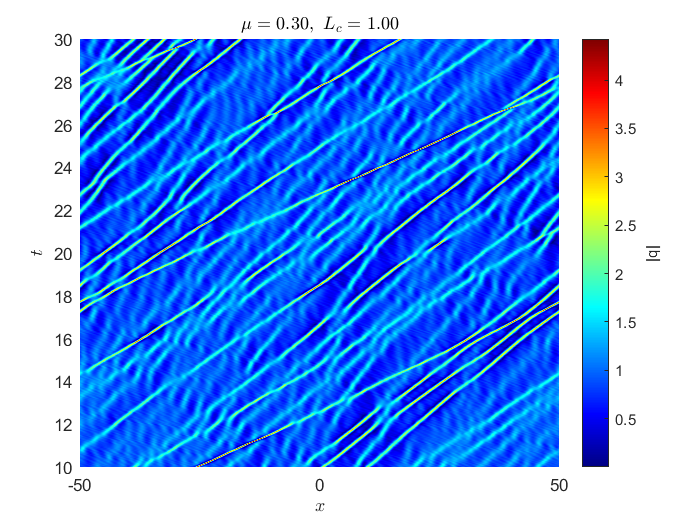}%
\par\smallskip
\noindent\parbox[t]{0.32\linewidth}{\centering\small\strut\textbf{(d)} $\mu=0.30$, $L_c=0.50$\par}%
\hfill%
\noindent\parbox[t]{0.32\linewidth}{\centering\small\strut\textbf{(e)} $\mu=0.30$, $L_c=0.70$\par}%
\hfill%
\noindent\parbox[t]{0.32\linewidth}{\centering\small\strut\textbf{(f)} $\mu=0.30$, $L_c=1.00$\par}%
\par
\caption{Two-dimensional spatiotemporal evolution of the chaotic wave field for different parameters.}
\label{fig:2d}
\end{figure}

Figure~\ref{fig:2d} shows the two-dimensional spatiotemporal evolution of the chaotic wave field for different parameters. The first row corresponds to fixed $L_c=0.50$ with $\mu = 0.10,0.30,0.60$ from left to right [Figs.~\ref{fig:2d}(a1)--(a3)], and the second row corresponds to fixed $\mu = 0.30$ with $L_c=0.50,0.70,1.00$ from left to right [Figs.~\ref{fig:2d}(b1)--(b3)].

It can be seen from the first row that a change in the standard deviation
\(\mu\) markedly alters the morphology of the chaotic wave field. For $\mu=0.10$
[Fig.~\ref{fig:2d}(a1)] only a few breather structures are scattered in the wave field, background radiation dominates, and the field is rather sparse. As $\mu$ increases to 0.30 and 0.60 [Figs.~\ref{fig:2d}(a2) and \ref{fig:2d}(a3)], the wave field is gradually filled with dense solitons, the number of bright spots increases appreciably, and the fluctuations become stronger. In addition, as can also be seen from the color scale, the larger \(\mu\) is, the higher the probability of high-amplitude values in the wave field, and the amplitude of some spikes already reaches several times that of the background wave field.

The second row illustrates the influence of the correlation length $L_c$ on the spatial structure of the chaotic wave field. Unlike the effect of \(\mu\), an increase in $L_c$ does not appreciably increase the number of bright spots, but rather widens the individual bright spots in space [Figs.~\ref{fig:2d}(b1)--(b3)]. This occurs because a larger $L_c$ implies a gentler spatial variation of the initial noise, and the long-wavelength components are more readily amplified in the early stage of the evolution. Hence, the resulting soliton and breather structures are larger in space. This indicates that $L_c$ mainly governs the spatial size of the structures in the wave field, whereas \(\mu\) influences more directly the density of the nonlinear structures and the overall complexity of the wave field.

\begin{figure}[H]
\centering
\noindent\includegraphics[width=0.32\linewidth,height=4.6cm,angle=0]{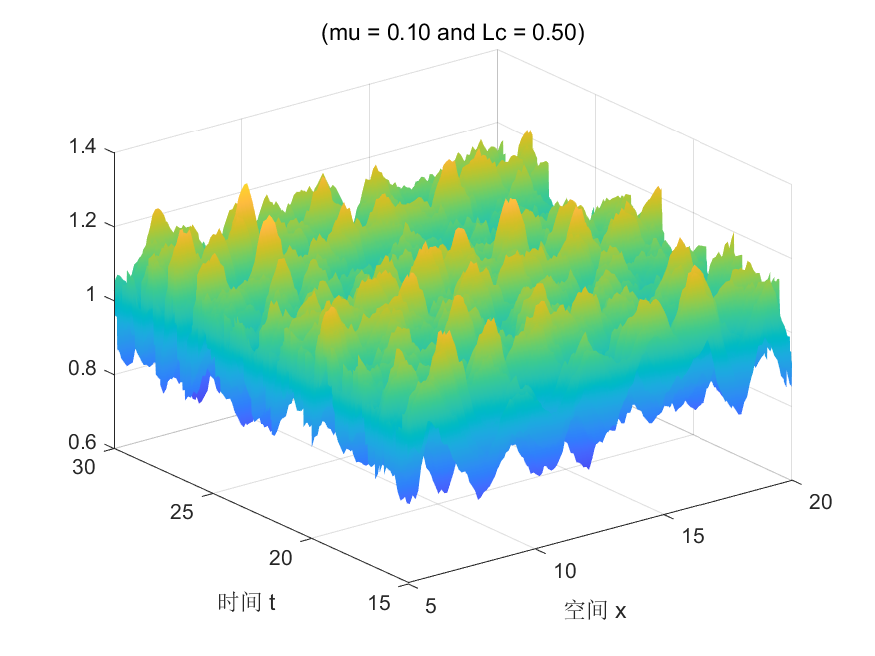}%
\hfill%
\noindent\includegraphics[width=0.32\linewidth,height=4.6cm,angle=0]{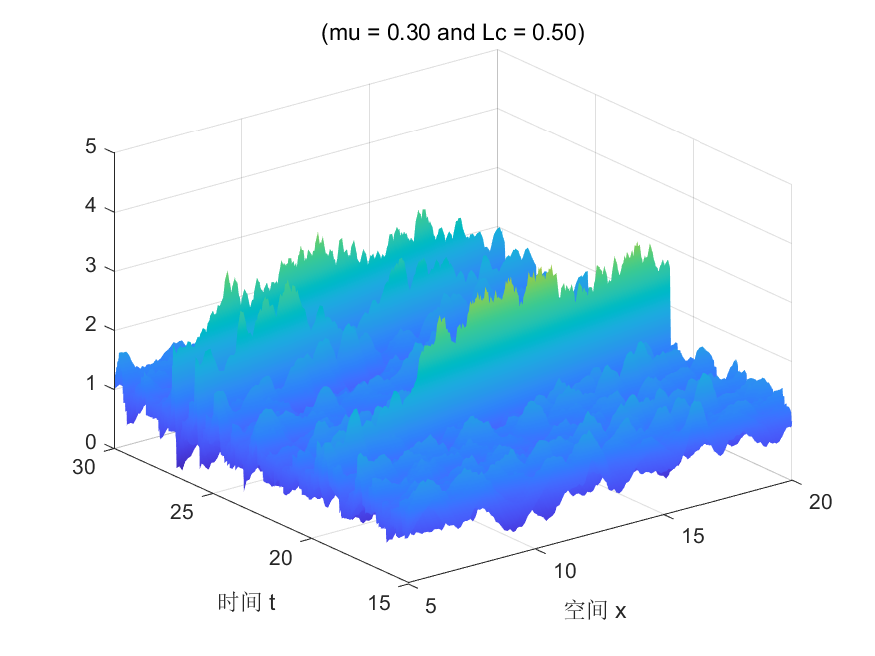}%
\hfill%
\noindent\includegraphics[width=0.32\linewidth,height=4.6cm,angle=0]{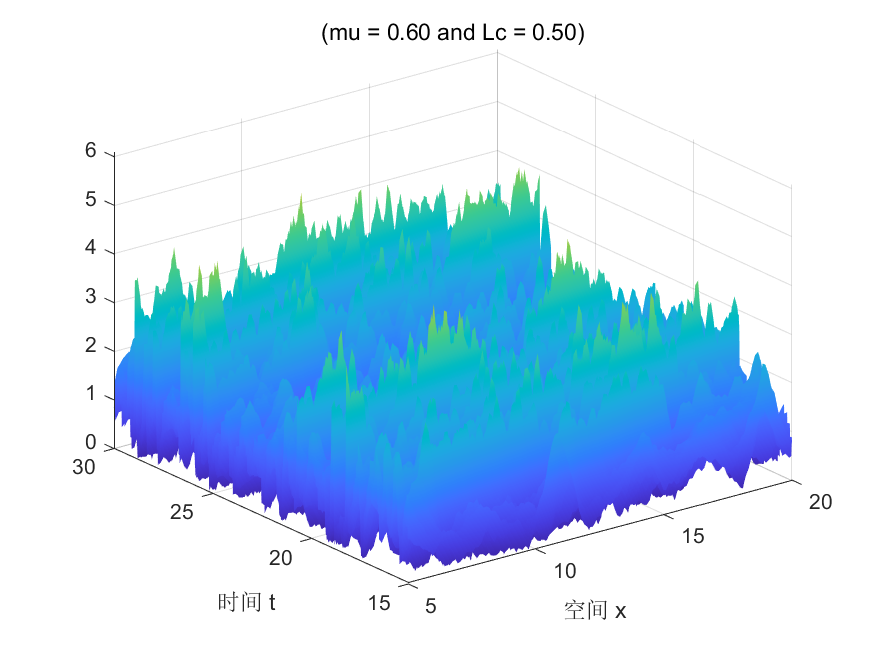}%
\par\smallskip
\noindent\parbox[t]{0.32\linewidth}{\centering\small\strut\textbf{(a)} $\mu=0.10$, $L_c=0.50$\par}%
\hfill%
\noindent\parbox[t]{0.32\linewidth}{\centering\small\strut\textbf{(b)} $\mu=0.30$, $L_c=0.50$\par}%
\hfill%
\noindent\parbox[t]{0.32\linewidth}{\centering\small\strut\textbf{(c)} $\mu=0.60$, $L_c=0.50$\par}%
\par
\medskip
\noindent\includegraphics[width=0.32\linewidth,height=4.6cm,angle=0]{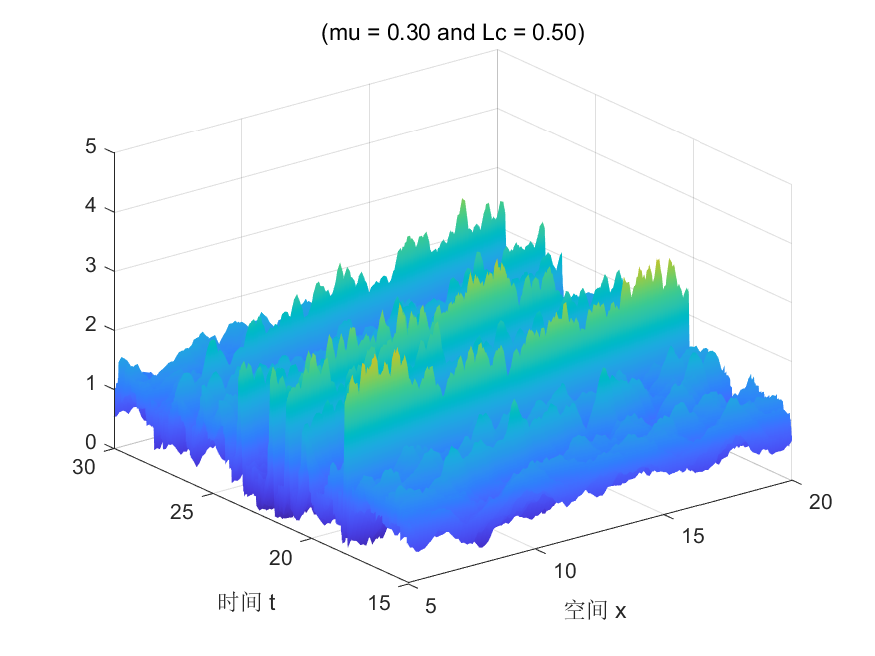}%
\hfill%
\noindent\includegraphics[width=0.32\linewidth,height=4.6cm,angle=0]{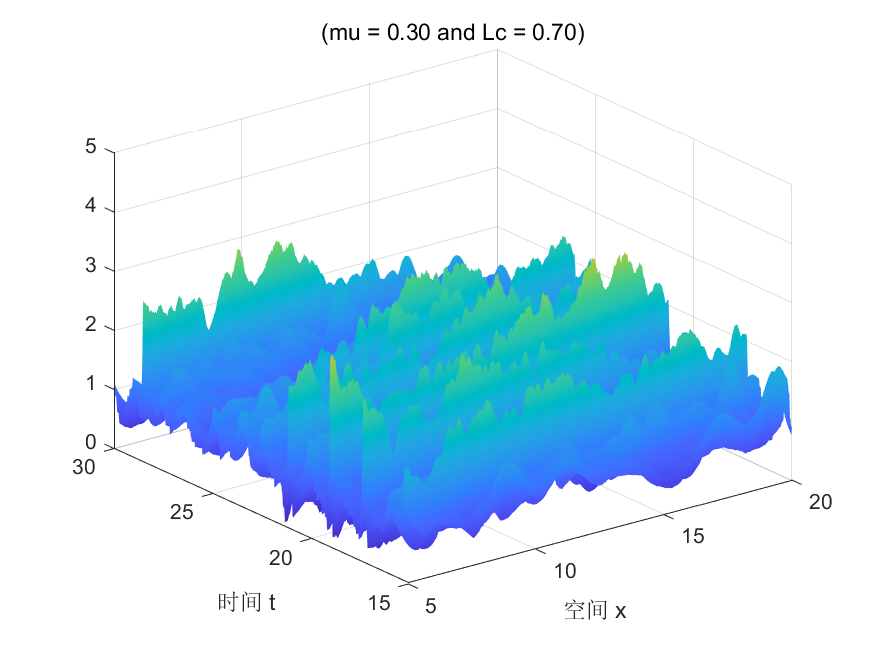}%
\hfill%
\noindent\includegraphics[width=0.32\linewidth,height=4.6cm,angle=0]{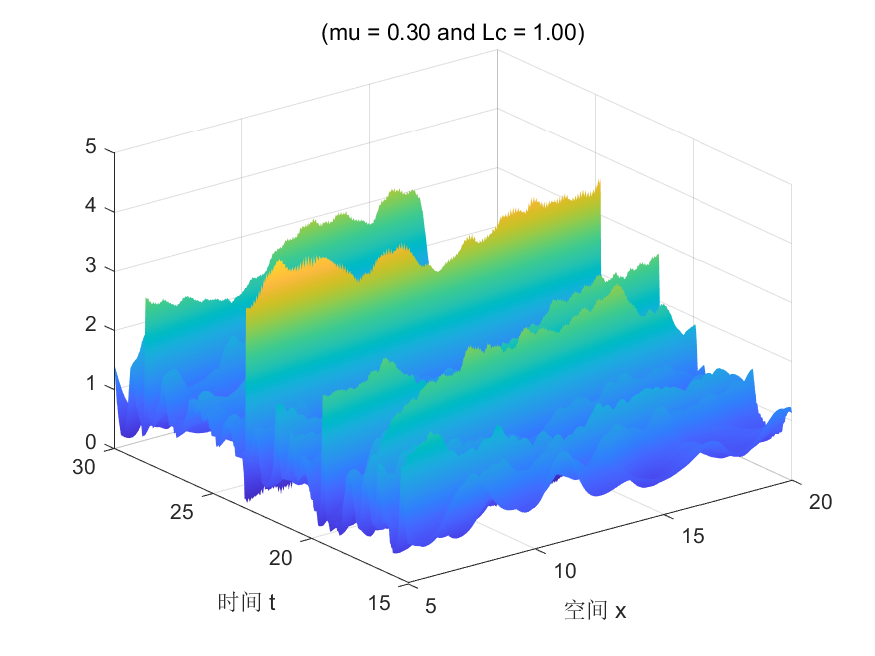}%
\par\smallskip
\noindent\parbox[t]{0.32\linewidth}{\centering\small\strut\textbf{(d)} $\mu=0.30$, $L_c=0.50$\par}%
\hfill%
\noindent\parbox[t]{0.32\linewidth}{\centering\small\strut\textbf{(e)} $\mu=0.30$, $L_c=0.70$\par}%
\hfill%
\noindent\parbox[t]{0.32\linewidth}{\centering\small\strut\textbf{(f)} $\mu=0.30$, $L_c=1.00$\par}%
\par
\caption{Three-dimensional spatiotemporal structure of the chaotic wave field for different parameters.}
\label{fig:3d}
\end{figure}

Figure~\ref{fig:3d} presents the evolution of the chaotic wave field from a three-dimensional perspective, where the $z$ axis denotes the wave field amplitude $|q|$, and the $(x,t)$ plane denotes the spatial and temporal coordinates. Compared with the two-dimensional top views, the three-dimensional plots reflect the amplitude fluctuations of the wave field more intuitively.

It can be seen from the first row that for $\mu=0.10$ [Fig.~\ref{fig:3d}(a1)] the wave field is rather flat, with only a few low peaks at isolated positions. As $\mu$ increases to 0.30 and 0.60 [Figs.~\ref{fig:3d}(a2) and \ref{fig:3d}(a3)], a large number of peak-like structures appear in the wave field, whose heights far exceed those of the surrounding background, and the higher the peak, the narrower and steeper the crest and the stronger the spatial localization. These spikes persist for only a short time and exhibit pronounced transient features, in accordance with the basic characteristics of rogue waves \cite{Akhmediev2009}.

The second row shows the three-dimensional structure for different $L_c$. Unlike the effect of increasing $\mu$, an increase in $L_c$ [Figs.~\ref{fig:3d}(b2) and \ref{fig:3d}(b3)] does not appreciably increase the height of the spikes, but broadens their bases in space so that the overall morphology becomes flatter. This contrast further confirms the conclusion drawn above that the correlation length $L_c$ mainly affects the spatial scale of the structures rather than their amplitude.

\subsection{Probability density function and rogue-wave generation probability}
\label{sec:pdf}

\begin{figure}[H]
\centering
\includegraphics[width=7.2cm,height=5.2cm,angle=0]{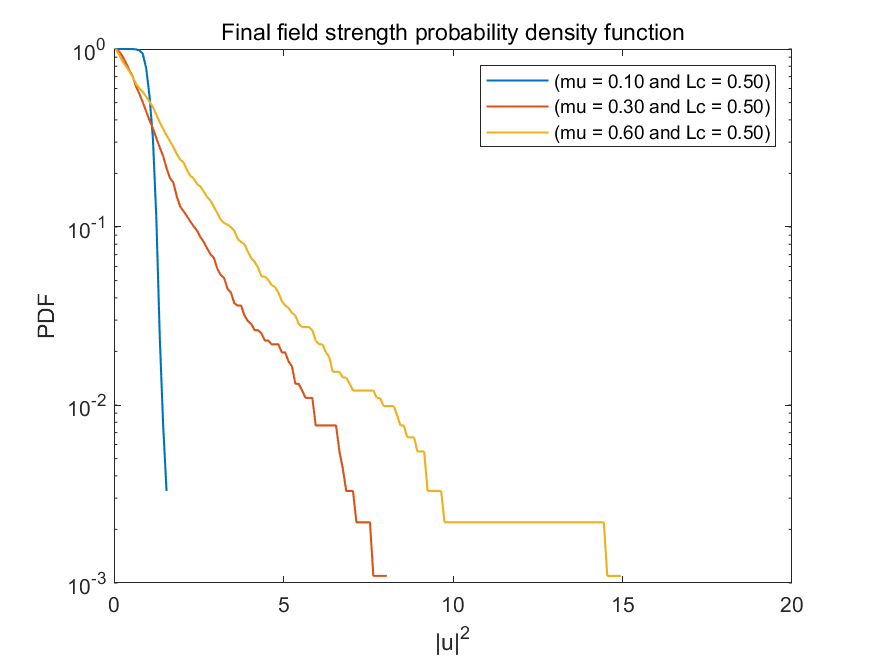}%
\hspace{0.3cm}%
\includegraphics[width=7.2cm,height=5.2cm,angle=0]{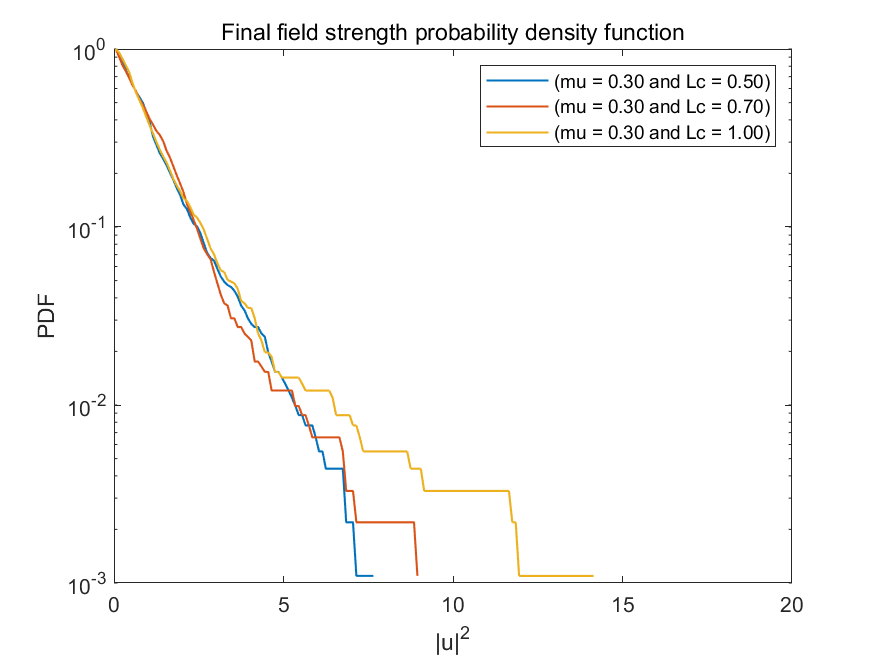}%
\par\smallskip
\parbox[t]{7.2cm}{\centering\small\strut\textbf{(a)} PDF for different $\mu$ ($L_c=0.5$)\par}%
\hspace{0.3cm}%
\parbox[t]{7.2cm}{\centering\small\strut\textbf{(b)} PDF for different $L_c$ ($\mu=0.30$)\par}%
\par
\caption{Probability density functions of the wave-field intensity in the late-time regime.}
\label{fig:pdf}
\end{figure}

To quantify the probability of rogue-wave occurrence, we computed the probability density function (PDF) of the wave-field intensity after the field fully evolved ($t=100$), as shown in Fig.~\ref{fig:pdf}. Figure~\ref{fig:pdf}(a) shows the PDFs for different values of $\mu$ at fixed $L_c=0.50$, with field intensity $|q|^2$ on the abscissa and logarithmic probability density on the ordinate. The PDF tail rises markedly as $\mu$ increases, indicating a substantial increase in the frequency of high-amplitude events. Since rogue waves are usually required to have a crest height exceeding twice that of the background field, this result shows that the likelihood of rogue-wave occurrence increases strongly with $\mu$. Figure~\ref{fig:pdf}(b) fixes $\mu=0.30$ and compares the PDFs for different $L_c$. Increasing $L_c$ also raises the PDF tail, but the effect is weaker than that of $\mu$. Together, the two panels show that $\mu$ is the dominant parameter controlling rogue-wave generation, while $L_c$ has a comparatively limited influence.

To quantify the frequency of rogue-wave events, we recorded the spatial distribution of the wave field every $\Delta t_s=0.5$ over $t\in[50,100]$, for a total of 101 records. For each record, we counted local maxima above the threshold (the rogue-wave count) and the fraction of spatial points above it (the rogue-wave fraction), then averaged both quantities over the 101 records. Because the field changes little between successive records, these records are not mutually independent. The vertical bar at each data point denotes the standard error of the mean, the uncertainty of the averaged value itself; regard a difference between two parameter points as statistically significant only if it exceeds this range.

\begin{figure}[H]
\centering
\includegraphics[width=7.2cm,height=5.2cm,angle=0]{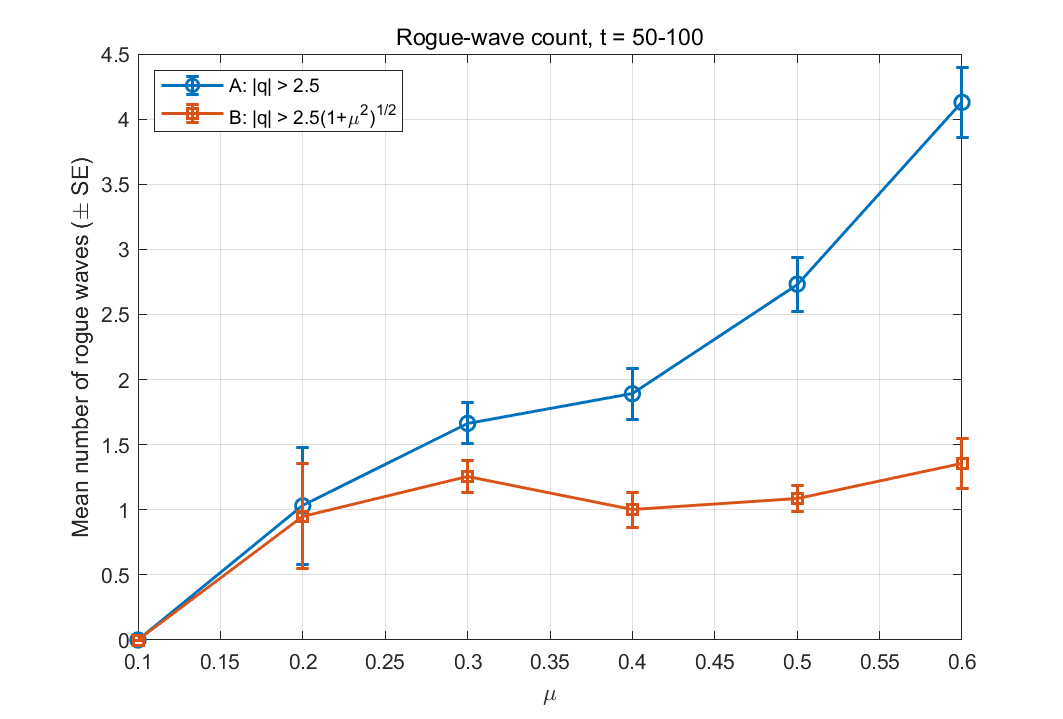}%
\hspace{0.3cm}%
\includegraphics[width=7.2cm,height=5.2cm,angle=0]{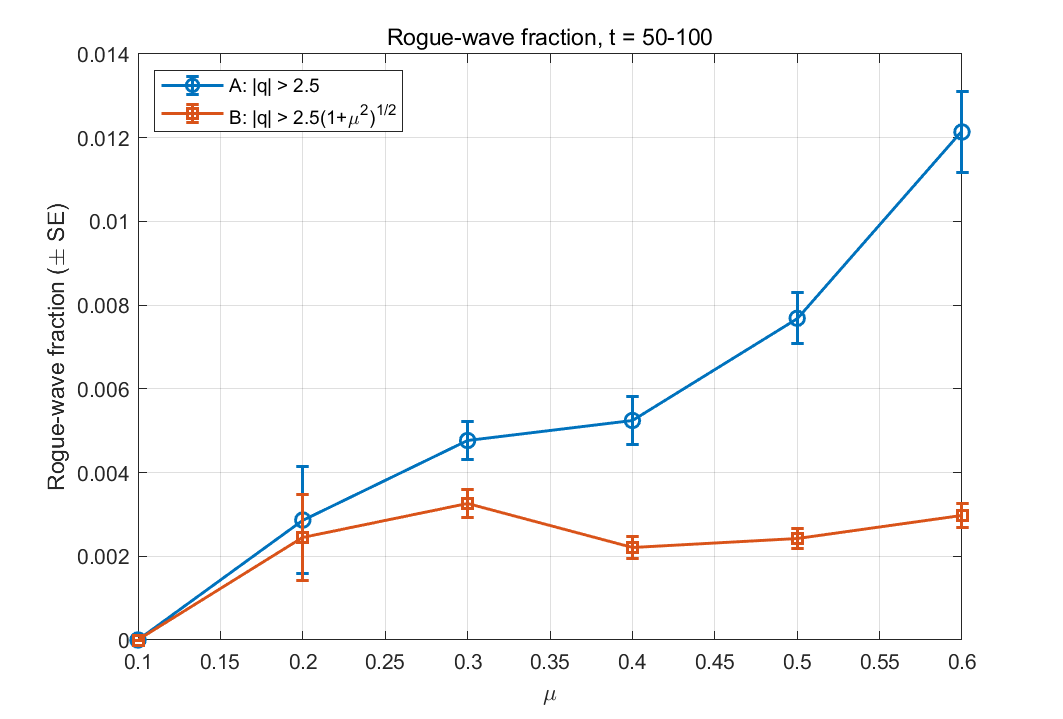}%
\par\smallskip
\parbox[t]{7.2cm}{\centering\small\strut\textbf{(a)} Rogue-wave count\par}%
\hspace{0.3cm}%
\parbox[t]{7.2cm}{\centering\small\strut\textbf{(b)} Rogue-wave fraction\par}%
\par
\caption{Time-averaged rogue-wave statistics for different initial perturbation strengths $\mu$ ($L_c=0.5$, observation window $t\in[50,100]$). Curves A and B correspond to fixed and amplitude-normalized thresholds, respectively.}
\label{fig:rogue_mu}
\end{figure}

Figure~\ref{fig:rogue_mu} shows how rogue-wave statistics vary with $\mu$ at fixed $L_c=0.5$. For the fixed threshold $|q|>2.5$, i.e., a crest height exceeding 2.5 times the background amplitude, the mean rogue-wave count increases monotonically from zero at $\mu=0.1$ to 4.13 at $\mu=0.6$ (curve A in Fig.~\ref{fig:rogue_mu}(a)), and the corresponding rogue-wave fraction rises from zero to $1.21\times10^{-2}$ (curve A in Fig.~\ref{fig:rogue_mu}(b)). The contrast between $\mu=0.2$ and $\mu=0.6$ is a factor of 4.2, corresponding to 5.8 standard errors. This trend agrees with the rise of the PDF tail with $\mu$ in Fig.~\ref{fig:pdf}(a). It shows that the rogue-wave generation probability increases significantly with the strength of the initial perturbation.

Since the amplitude of the random perturbation in the initial field $q(x,0)=1+\mu f(x)$ is proportional to $\mu$, a larger $\mu$ produces larger overall field fluctuations. Consequently, the fixed threshold $|q|>2.5$ becomes increasingly low relative to the field, so more points exceed it. To isolate this effect, we adopted a second threshold, $|q|>2.5\sqrt{1+\mu^2}$, which also increases with $\mu$ and equals 2.5 times the root-mean-square amplitude of the initial field (curve B in Fig.~\ref{fig:rogue_mu}(a,b)). With this threshold, the mean rogue-wave count remains between 0.95 and 1.36, and the rogue-wave fraction between $2.2\times10^{-3}$ and $3.3\times10^{-3}$, with no systematic dependence on $\mu$. This indicates that, at a fixed threshold, the increase in rogue-wave probability comes mainly from the larger amplitude of the initial perturbation rather than from additional nonlinear amplification; that is, the field generates extreme events approximately in proportion to the strength of the initial perturbation. For small $\mu$, $\sqrt{1+\mu^2}\approx1$, so the two curves almost coincide, whereas for large $\mu$ they separate appreciably.

Furthermore, the error range at $\mu=0.2$ is notably larger than at the other points. At this value, strange waves appear intermittently: long periods show no pronounced large-amplitude structures, followed by several rogue waves in a short interval. This increases fluctuations in the statistic over time and, in turn, the standard error. For $\mu\ge0.3$, rogue waves appear more persistently, and the standard error decreases accordingly. A similar behavior is observed when $L_c$ is varied.

\begin{figure}[H]
\centering
\includegraphics[width=7.2cm,height=5.2cm,angle=0]{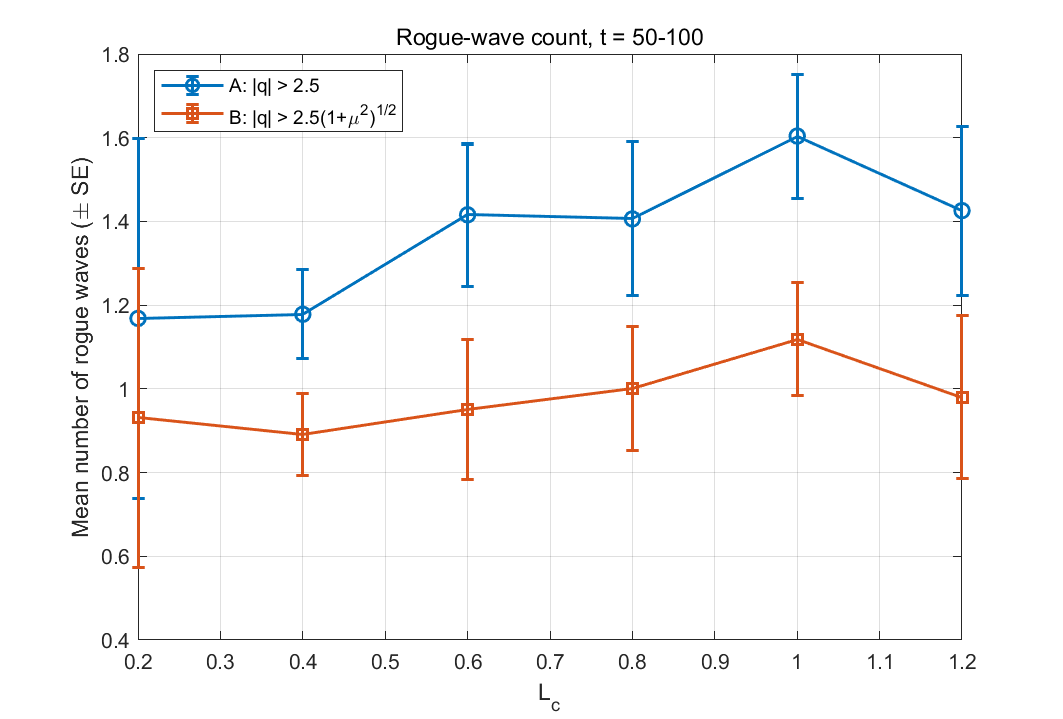}%
\hspace{0.3cm}%
\includegraphics[width=7.2cm,height=5.2cm,angle=0]{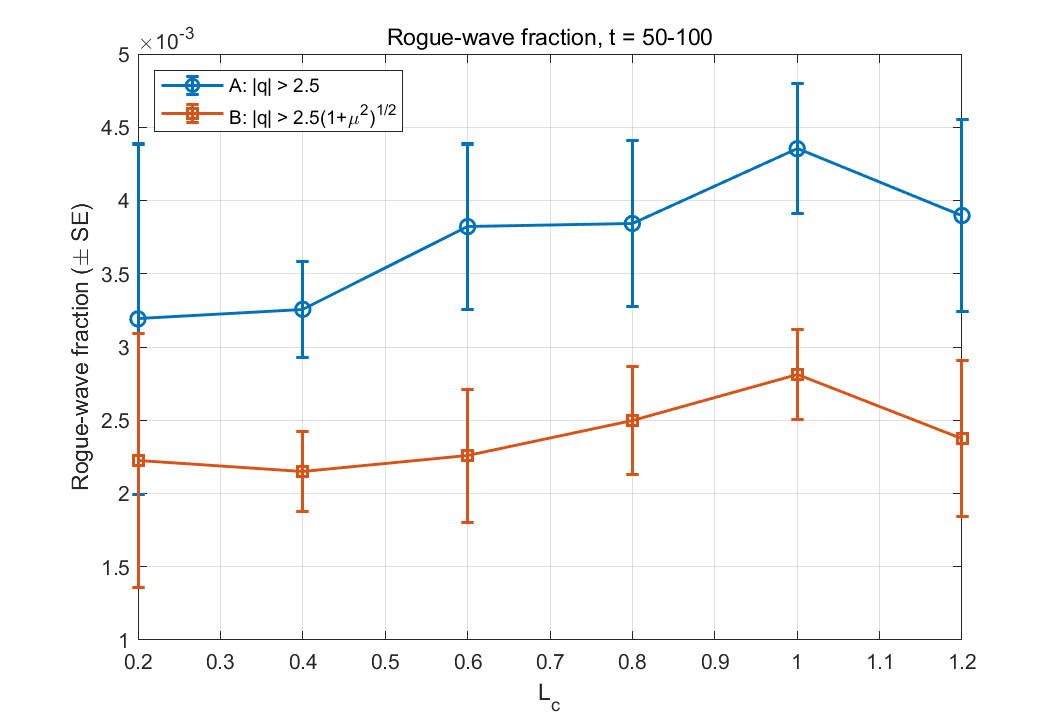}%
\par\smallskip
\parbox[t]{7.2cm}{\centering\small\strut\textbf{(a)} Rogue-wave count\par}%
\hspace{0.3cm}%
\parbox[t]{7.2cm}{\centering\small\strut\textbf{(b)} Rogue-wave fraction\par}%
\par
\caption{Time-averaged rogue-wave statistics for different correlation lengths $L_c$ ($\mu=0.3$, $t\in[50,100]$). Curves A and B are defined as in Fig.~\ref{fig:rogue_mu}.}
\label{fig:rogue_Lc}
\end{figure}

Figure~\ref{fig:rogue_Lc} shows the results obtained by varying $L_c$ at fixed $\mu=0.3$. Over $L_c\in[0.2,1.2]$, the average rogue-wave count lies between 1.17 and 1.60, and the rogue-wave fraction between $3.2\times10^{-3}$ and $4.4\times10^{-3}$; all differences between data points are smaller than the corresponding standard errors. Therefore, $L_c$ has no significant influence on the probability of rogue-wave generation. This result is consistent with the slight rise of the PDF tail in Fig.~\ref{fig:pdf}(b) and confirms the conclusion of Sec.~3.2 that $L_c$ mainly governs the spatial scale of wave-field structures without altering the frequency of high-amplitude events.

\subsection{Eigenvalue spectrum analysis and transition between turbulence types}
\label{sec:eigen}

The PDF analysis shows that an increase in the standard deviation $\mu$ significantly enhances the probability of rogue wave generation. This increase also indicates a transition in turbulence type, which is further revealed by the Lax pair eigenvalue spectrum in this subsection.

The analysis in Section~\ref{sec:spectral} shows that the different nonlinear structures have well-defined counterparts on the spectral plane of the Lax pair. Specifically, solitons correspond to complex-conjugate eigenvalues with nonzero real part, and breathers correspond to eigenvalues on the imaginary axis. Therefore, by counting the distribution of the eigenvalues of the chaotic wave field on the complex plane, the relative proportion of solitons and breathers in the wave field can be determined, and the type of turbulence can be identified \cite{SotoCrespo2016, Akhmediev2016, El2020}.

\begin{figure}[H]
\centering
\noindent\includegraphics[width=0.32\linewidth,height=4.6cm,angle=0]{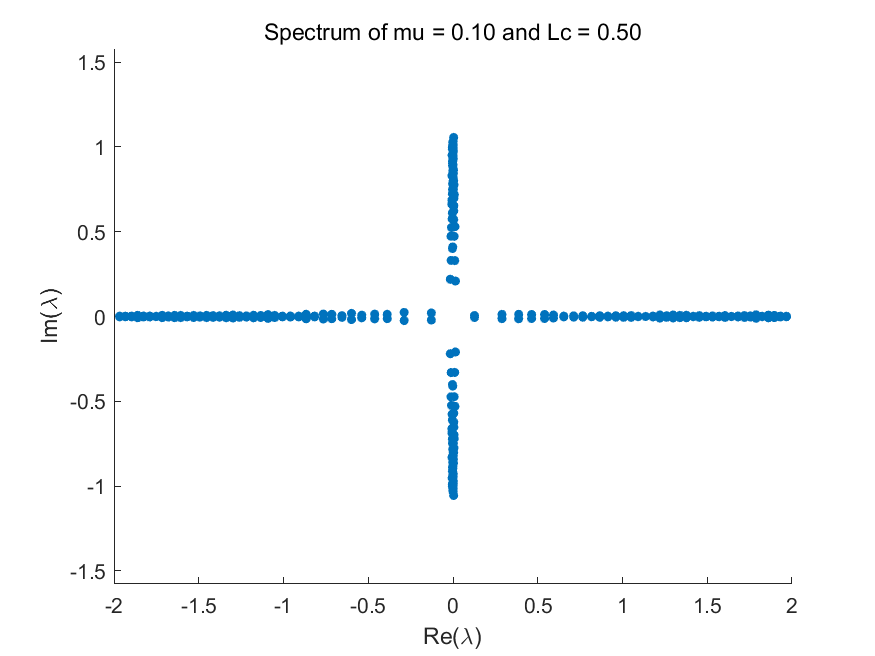}%
\hfill%
\noindent\includegraphics[width=0.32\linewidth,height=4.6cm,angle=0]{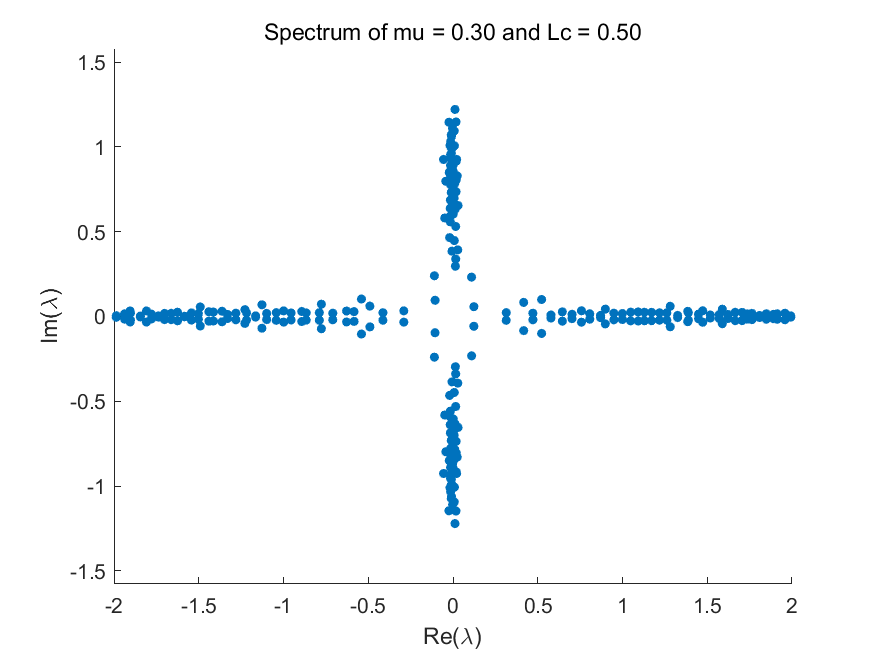}%
\hfill%
\noindent\includegraphics[width=0.32\linewidth,height=4.6cm,angle=0]{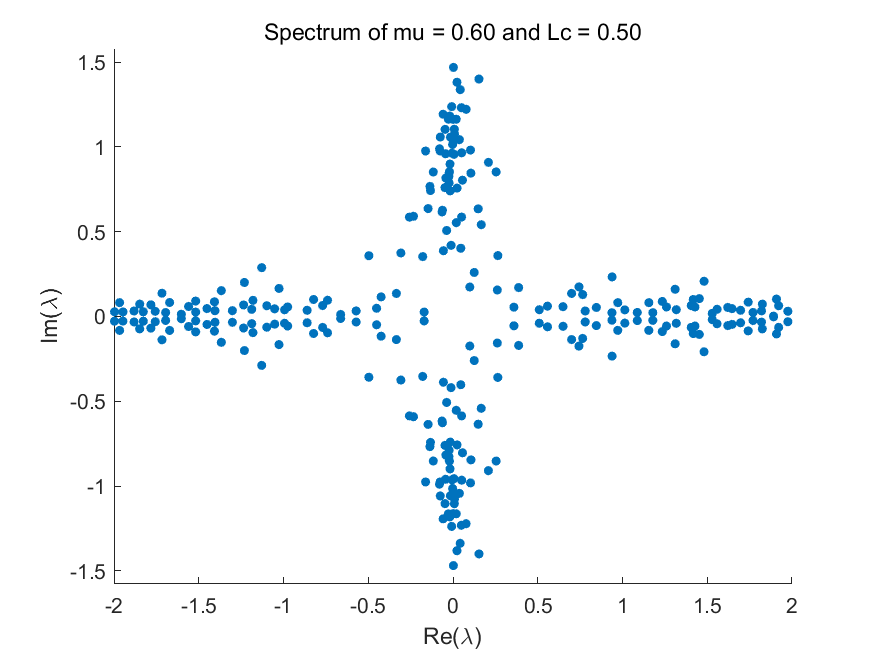}%
\par\smallskip
\noindent\parbox[t]{0.32\linewidth}{\centering\small\strut\textbf{(a)} $\mu=0.10$\par}%
\hfill%
\noindent\parbox[t]{0.32\linewidth}{\centering\small\strut\textbf{(b)} $\mu=0.30$\par}%
\hfill%
\noindent\parbox[t]{0.32\linewidth}{\centering\small\strut\textbf{(c)} $\mu=0.60$\par}%
\par
\medskip
\noindent\includegraphics[width=0.32\linewidth,height=4.6cm,angle=0]{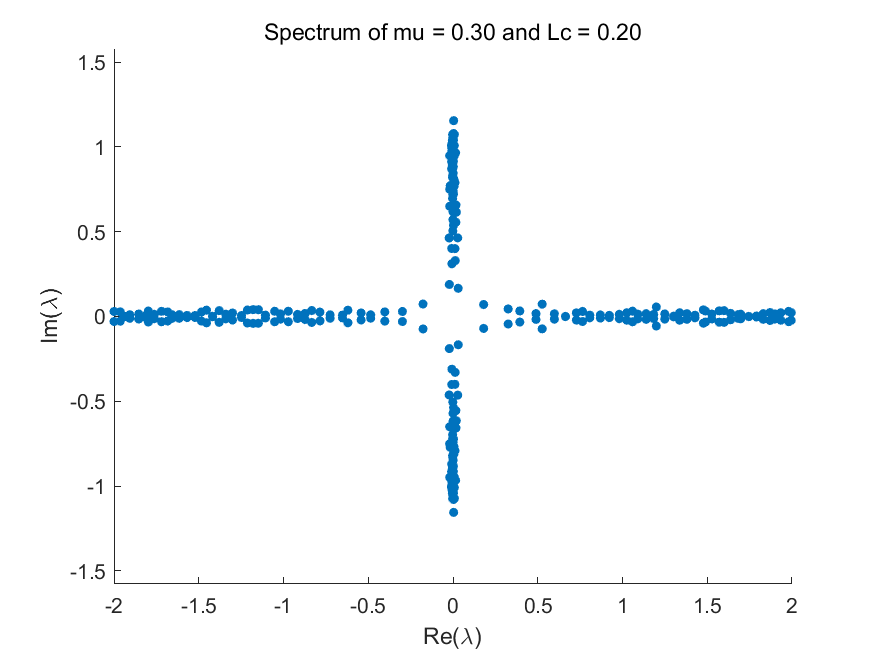}%
\hfill%
\noindent\includegraphics[width=0.32\linewidth,height=4.6cm,angle=0]{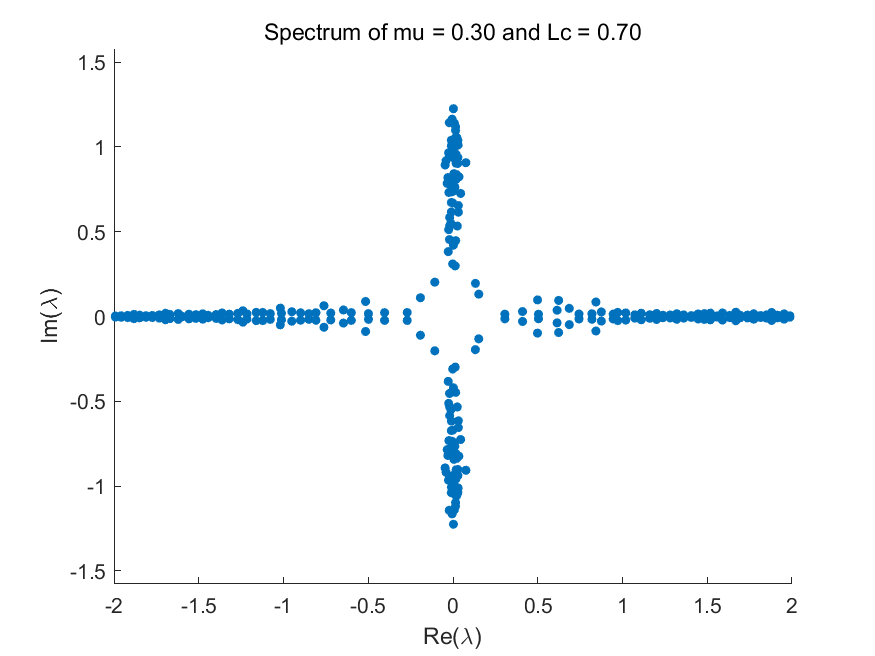}%
\hfill%
\noindent\includegraphics[width=0.32\linewidth,height=4.6cm,angle=0]{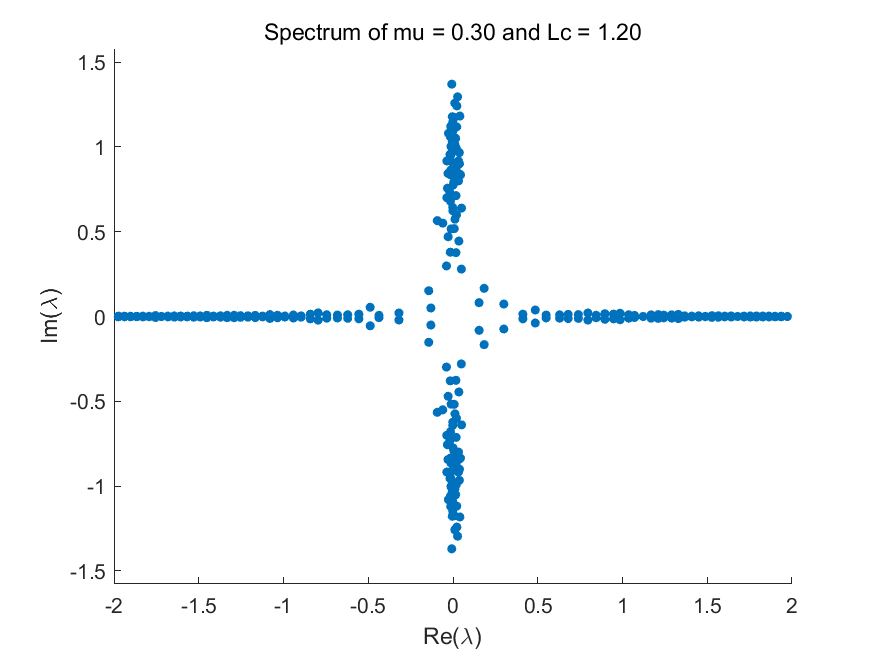}%
\par\smallskip
\noindent\parbox[t]{0.32\linewidth}{\centering\small\strut\textbf{(d)} $L_c=0.50$\par}%
\hfill%
\noindent\parbox[t]{0.32\linewidth}{\centering\small\strut\textbf{(e)} $L_c=0.70$\par}%
\hfill%
\noindent\parbox[t]{0.32\linewidth}{\centering\small\strut\textbf{(f)} $L_c=1.00$\par}%
\par
\caption{Lax-pair eigenvalue spectra of the chaotic wave field for different parameters. In the first row, $L_c=0.50$ is fixed and $\mu=0.10,0.30,0.60$ from left to right; in the second row, $\mu=0.30$ is fixed and $L_c=0.50,0.70,1.00$ from left to right.}
\label{fig:spectra}
\end{figure}

\begin{figure}[H]
\centering
\includegraphics[width=7.2cm,height=5.2cm,angle=0]{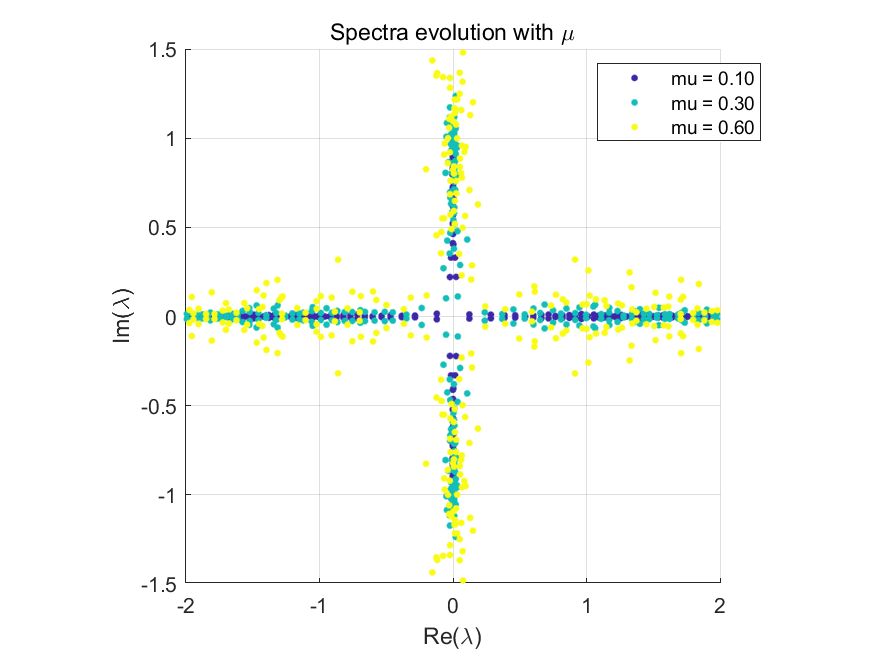}%
\hspace{0.3cm}%
\includegraphics[width=7.2cm,height=5.2cm,angle=0]{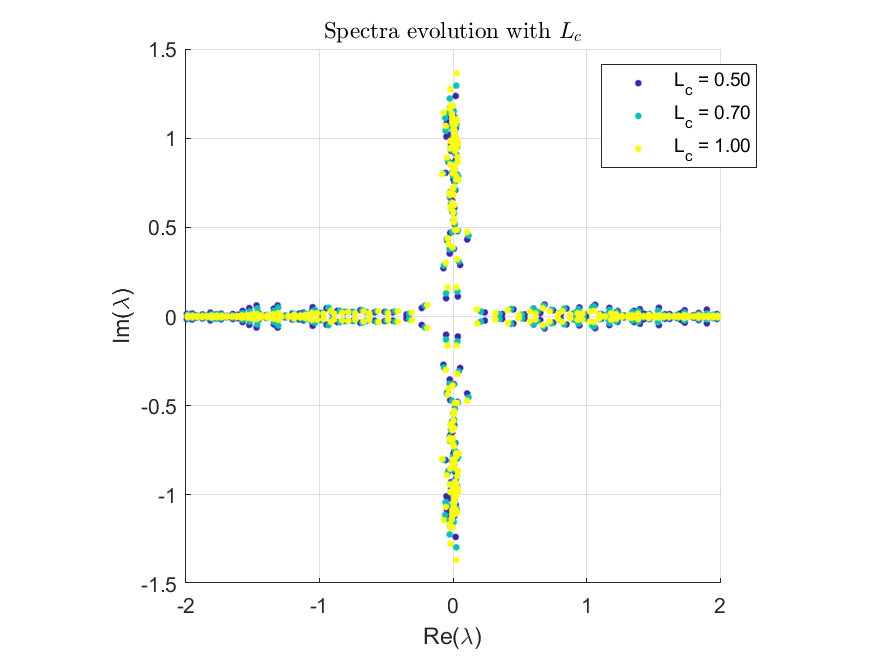}%
\par\smallskip
\parbox[t]{7.2cm}{\centering\small\strut\textbf{(a)} Spectral comparison for different $\mu$ ($L_c=0.50$)\par}%
\hspace{0.3cm}%
\parbox[t]{7.2cm}{\centering\small\strut\textbf{(b)} Spectral comparison for different $L_c$ ($\mu=0.30$)\par}%
\par
\caption{Superimposed comparison of the eigenvalue spectra for different parameters.}
\label{fig:spectra_compare}
\end{figure}

Since the spectral parameter remains invariant during the evolution (see Section~\ref{sec:spectral}), the spectra shown in Figs.~\ref{fig:spectra} and \ref{fig:spectra_compare} can be computed directly from the initial field $q(x,0)$. Figures~\ref{fig:spectra}(a1)--(a3) present the eigenvalue spectra of the Lax pair for different $\mu$ at the fixed value $L_c=0.5$. For $\mu=0.10$ [Fig.~\ref{fig:spectra}(a1)], the eigenvalues are distributed along the imaginary and real axes and include breathers and background radiation, with almost no solitons, so that breather turbulence is dominant. As $\mu$ increases to $0.30$ [Fig.~\ref{fig:spectra}(a2)], some eigenvalues begin to deviate from the imaginary axis, and the number of points with nonzero real part gradually increases, indicating the appearance of a soliton component. When $\mu$ is further increased to $0.60$ [Fig.~\ref{fig:spectra}(a3)], the distribution range of the eigenvalues on the complex plane is markedly enlarged, and a large number of eigenvalues possess a pronounced real part, which means that solitons have become the main constituent of the wave field and that the turbulence type has changed from breather turbulence to soliton turbulence.
Figure~\ref{fig:spectra_compare}(a) superimposes the three cases, making the gradual process even clearer and showing that, with increasing $\mu$, the eigenvalue distribution spreads gradually from the vicinity of the imaginary axis towards both sides of the complex plane.

The change in the spectral distribution agrees with the observations from the two-dimensional evolution plots in Section~\ref{sec:evolution}. For small $\mu$, the wave field is sparse and dominated by breathers; for larger $\mu$, solitons appear in large numbers, and the wave field becomes progressively denser. This transition in the eigenvalue spectrum, from concentration along the imaginary axis to spread across the complex plane, provides strong evidence for the shift in turbulence type from breather-dominated to soliton-dominated. \cite{Akhmediev2016}.

Figures~\ref{fig:spectra}(b1)--(b3) show the eigenvalue spectra for different $L_c$ at fixed $\mu=0.30$. It can be seen that increasing $L_c$ does not cause the eigenvalues to deviate significantly from the imaginary axis; in all three panels, the eigenvalues are mainly distributed in the vicinity of the imaginary axis, and the proportion of eigenvalues with nonzero real part changes little. This indicates that changing $L_c$ mainly affects the spatial scale of the structures in the wave field (as shown in Fig.~\ref{fig:2d}), but does not alter the relative proportion of solitons and breathers, and hence cannot induce an intrinsic transition of the turbulence type. Figure~\ref{fig:spectra_compare}(b) likewise superimposes the spectra for the three values of $L_c$; although the distribution range of the spectrum expands slightly as $L_c$ increases, the extent is far less pronounced than that caused by a change of $\mu$.

From the analysis of Figs.~\ref{fig:spectra} and \ref{fig:spectra_compare}, the following conclusion can be drawn: the standard deviation $\mu$ is the key parameter controlling the transition of the integrable turbulence type in the complex mKdV equation---an increase in $\mu$ drives the system from breather turbulence to soliton turbulence, whereas the correlation length $L_c$ mainly affects the spatial scale of the structures and has only a weak influence on the turbulence type. Since extreme events in soliton-dominated turbulence originate mainly from strong interactions between solitons \cite{Zhang2024SolitonGas, Zhang2024Gardner}, this also explains why the probability of large-amplitude events increases significantly for large $\mu$. This conclusion is consistent with the results obtained for the KE equation \cite{Li2024KE}.

To quantify the transition described above, we counted the spectral points. The continuous spectrum of the constant background consists of the imaginary segment $[-ic, ic]$ and the real axis, excluding the isolated points; points lying on it correspond to background radiation and are therefore excluded from the statistics. The remaining isolated points are classified according to Sec.~2: those with a nonzero real part are counted as solitons, and those on the imaginary axis as breathers. Note that for $0<\beta<c$ (spatial breathers) the eigenvalues $\pm i\beta$ are embedded in the continuous spectrum and cannot be distinguished from it numerically; these eigenvalues are therefore not included, so that the breathers counted below correspond to the temporal breathers with $\beta>c$.

\begin{figure}[H]
\centering
\includegraphics[width=7.2cm,height=5.2cm,angle=0]{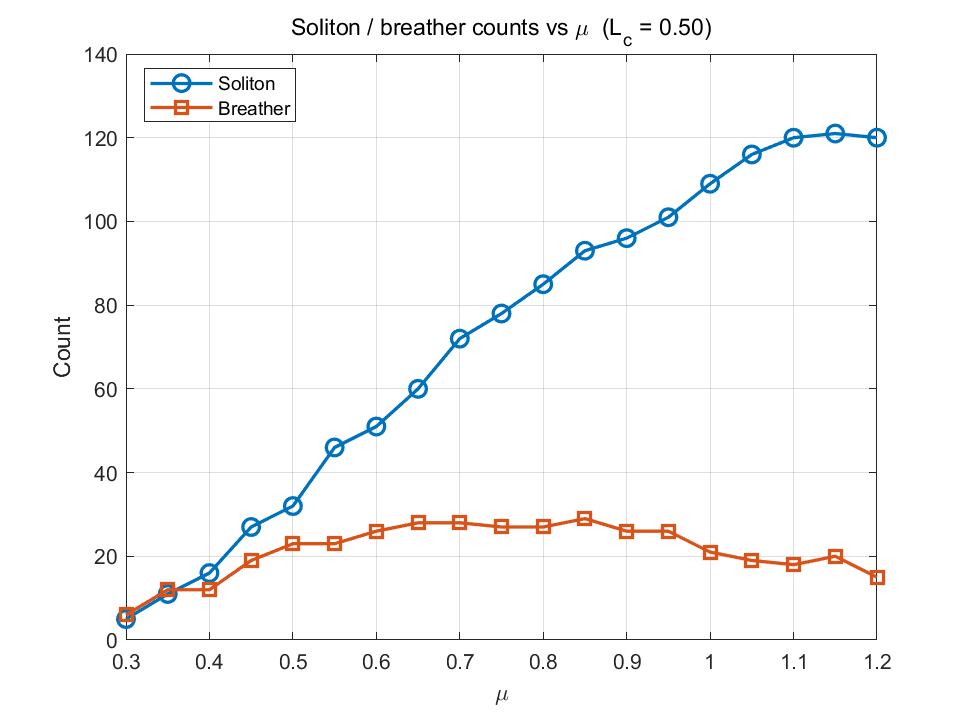}%
\hspace{0.3cm}%
\includegraphics[width=7.2cm,height=5.2cm,angle=0]{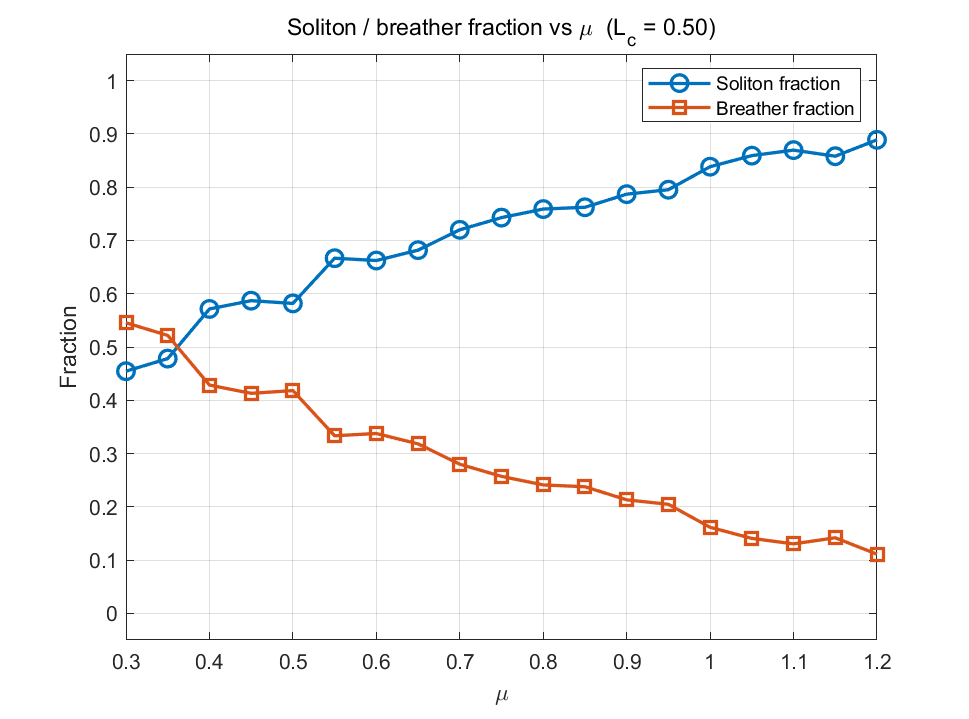}%
\par\smallskip
\parbox[t]{7.2cm}{\centering\small\strut\textbf{(a)} Count\par}%
\hspace{0.3cm}%
\parbox[t]{7.2cm}{\centering\small\strut\textbf{(b)} Fraction\par}%
\par
\caption{Numbers and fractions of solitons and breathers for different initial perturbation strengths $\mu$ ($L_c=0.5$).}
\label{fig:sol_bre_mu}
\end{figure}

Figure~\ref{fig:sol_bre_mu} shows the numbers and fractions of solitons and breathers versus $\mu$ at fixed $L_c=0.5$. The soliton count increases monotonically from 5 at $\mu=0.30$ to 120 at $\mu=1.20$, while the breather count first increases from 6 to about 29 and then decreases to 15. Correspondingly, the soliton fraction rises monotonically from 0.455 to 0.889 and crosses 0.5 near $\mu\approx0.40$. Thus, for small $\mu$ the numbers of solitons and breathers are comparable and the turbulence is in a mixed state; as $\mu$ increases, solitons multiply rapidly and become dominant, and the turbulence changes from breather-dominated to soliton-dominated. This result is consistent with the spreading of the eigenvalues towards both sides of the complex plane in Fig.~\ref{fig:spectra}, and with the evolution of the field from sparse to dense described in Sec.~3.2.

\begin{figure}[H]
\centering
\includegraphics[width=7.2cm,height=5.2cm,angle=0]{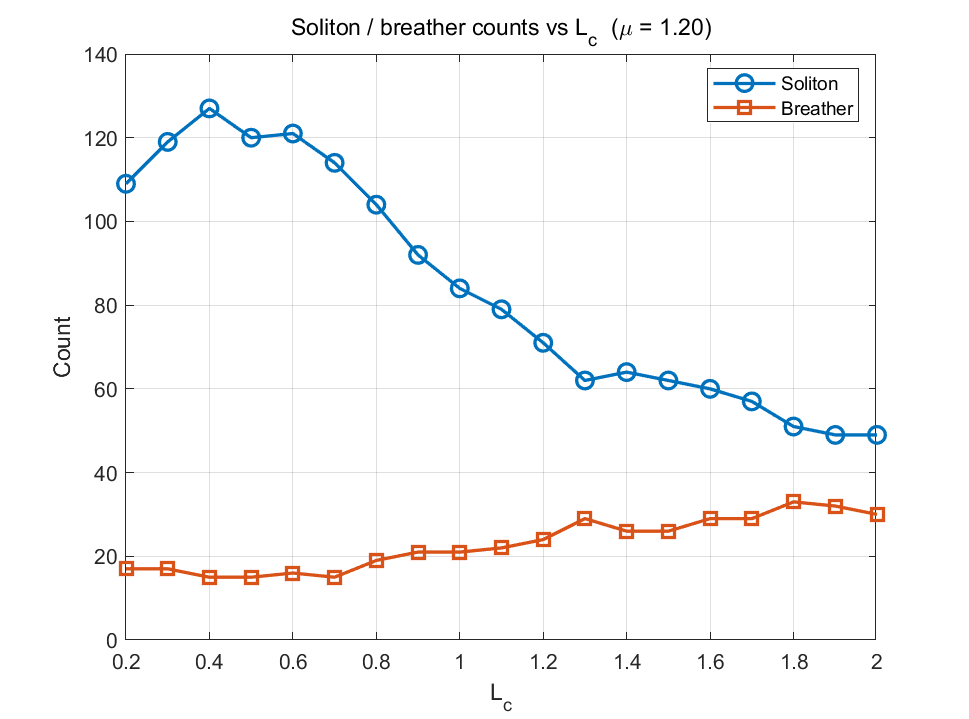}%
\hspace{0.3cm}%
\includegraphics[width=7.2cm,height=5.2cm,angle=0]{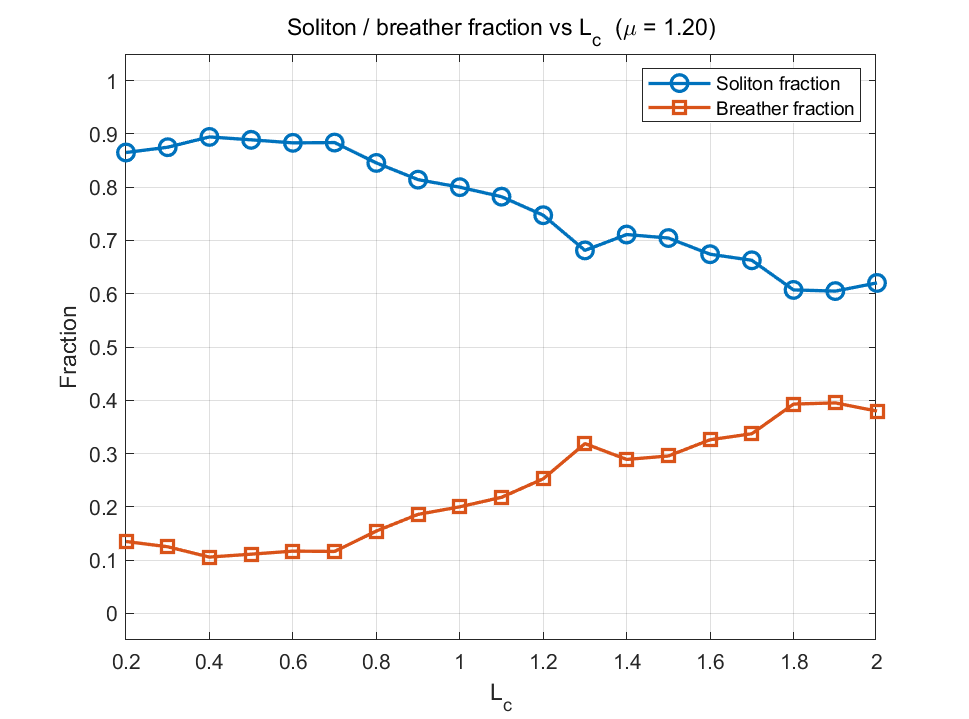}%
\par\smallskip
\parbox[t]{7.2cm}{\centering\small\strut\textbf{(a)} Count\par}%
\hspace{0.3cm}%
\parbox[t]{7.2cm}{\centering\small\strut\textbf{(b)} Fraction\par}%
\par
\caption{Numbers and fractions of solitons and breathers for different correlation lengths $L_c$ ($\mu=1.2$).}
\label{fig:sol_bre_Lc}
\end{figure}

Figure~\ref{fig:sol_bre_Lc} shows soliton and breather statistics versus $L_c$ under strong perturbation ($\mu=1.2$). As $L_c$ increases from 0.2 to 2.0, the soliton count decreases from 109 to 49 while the breather count increases from 17 to 30, and the soliton fraction decreases correspondingly from 0.865 to 0.620. Thus $L_c$ also modifies the relative proportions of solitons and breathers appreciably, but in a different manner from $\mu$. Over the whole range of $L_c$, the soliton fraction remains above 0.6, so solitons always dominate, and $L_c$ merely adjusts the relative numbers of the two species. This behavior is consistent with the conclusion of Sec.~3.2 that $L_c$ determines the spatial scale of the field structures. Consequently, for larger $L_c$, the initial perturbation varies more gently in space, and the resulting structures are wider and sparser, so the number of solitons decreases and the fraction of breathers increases.

Figures~\ref{fig:sol_bre_mu} and \ref{fig:sol_bre_Lc} show 121 and 120 solitons at the common parameter point $(\mu=1.2,\ L_c=0.5)$, respectively. The agreement between the two values confirms the consistency of the statistics.

In summary, the standard deviation $\mu$ is the key parameter controlling the transition between turbulence types in the complex mKdV equation: increasing $\mu$ drives the system from breather turbulence to soliton turbulence. The correlation length $L_c$ mainly determines the spatial scale of the field structures; it changes the relative proportions of solitons and breathers, but does not alter the dominance of solitons. Since extreme events in soliton-dominated turbulence originate mainly from strong interactions between solitons~\cite{Zhang2024SolitonGas, Zhang2024Gardner}, this transition from breather to soliton dominance also explains the pronounced increase in the rogue-wave generation probability with $\mu$ reported in Sec.~3.3. This conclusion is consistent with the results obtained for the KE equation~\cite{Li2024KE}.

\subsection{Energy evolution in the wavenumber domain and wave-action spectrum analysis}
\label{sec:waveaction}

The preceding sections analyzed the evolution of the chaotic wave field in physical space, i.e., in the $(x,t)$ domain. To further investigate the transfer process of wave field energy in the frequency domain, this section applies a Fourier transform to the wave field $q(x,t)$ and calculates its wave action spectrum $S_k(t)$ \cite{Agafontsev2015}. Compared to the spatial domain, this frequency-domain analysis provides a more intuitive illustration of how nonlinear evolution selectively amplifies perturbations at specific scales.

The wave-action spectrum is defined as
\begin{equation}
S_k(t) = |\hat{q}_k(t)|^2,
\label{eq:waveaction}
\end{equation}
where $\hat{q}_k(t)$ is the discrete Fourier transform of $q(x,t)$, namely
\begin{equation}
\hat{q}_k(t) = \mathcal{F}[q(x,t)]
= \frac{1}{L}\int_{-L/2}^{L/2} q(x,t)e^{-ikx}dx,
\label{eq:fourier}
\end{equation}
with $k = 2\pi n/L$ the wave number (where $n$ is an integer) and $L$ the length of the spatial interval. The wave action spectrum describes the distribution of energy across different wave numbers, and it satisfies the following relationship with the total wave action $N$.
\begin{equation}
N = \sum_k S_k(t).
\label{eq:totalaction}
\end{equation}

\begin{figure}[H]
\centering
\noindent\includegraphics[width=0.32\linewidth,height=4.6cm,angle=0]{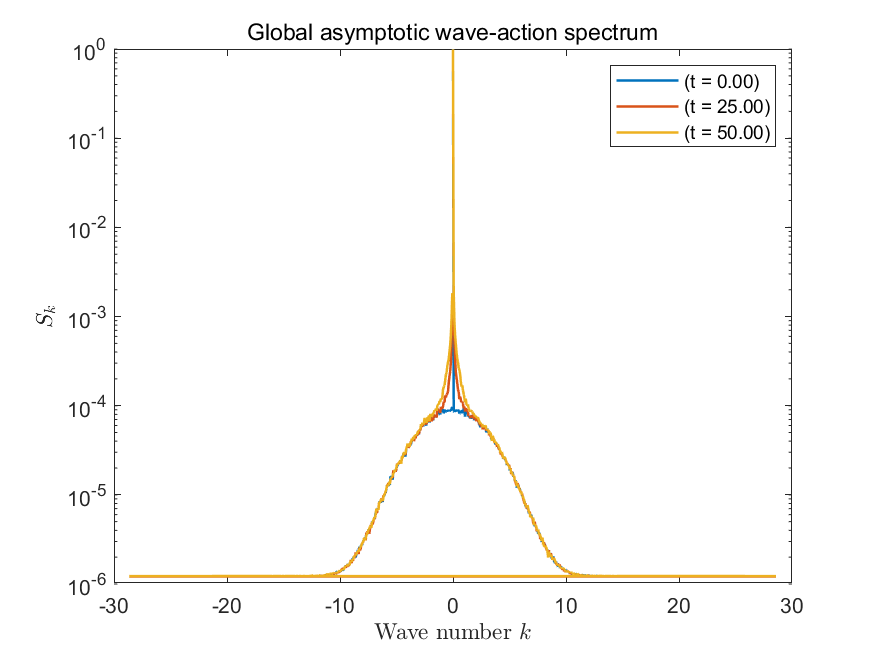}%
\hfill%
\noindent\includegraphics[width=0.32\linewidth,height=4.6cm,angle=0]{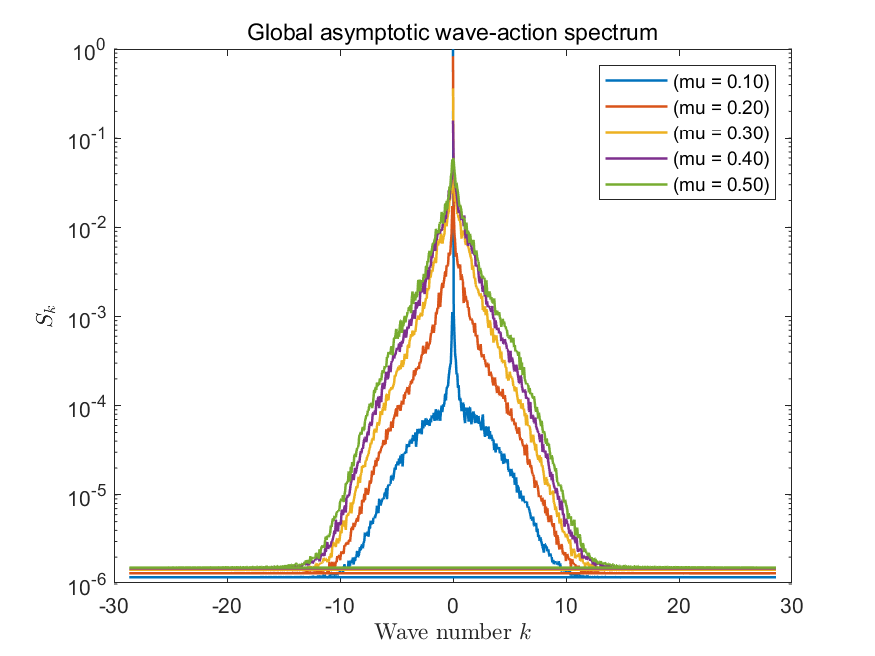}%
\hfill%
\noindent\includegraphics[width=0.32\linewidth,height=4.6cm,angle=0]{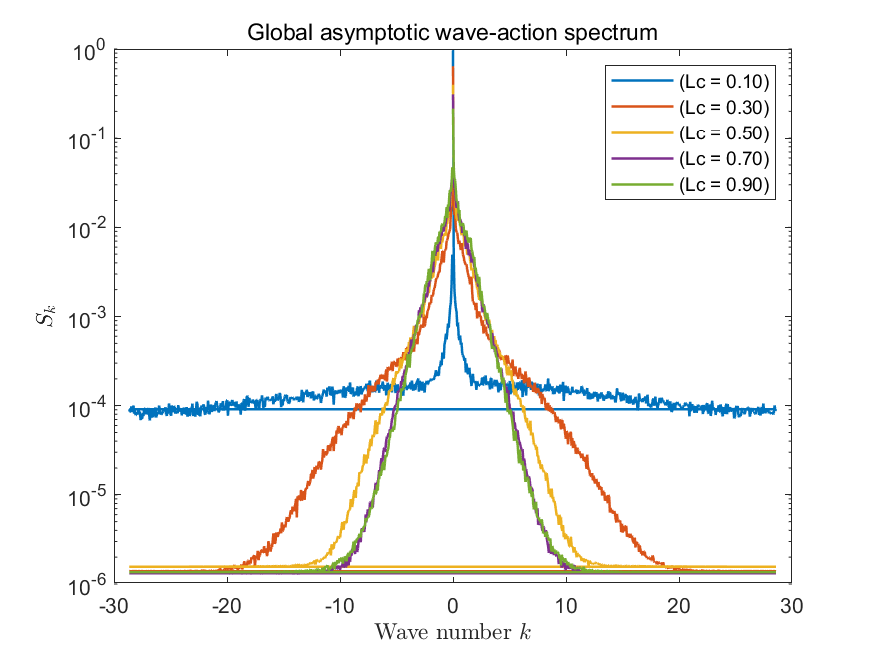}%
\par\smallskip
\noindent\parbox[t]{0.32\linewidth}{\centering\small\strut\textbf{(a)} Evolution of the wave-action spectrum with time\par}%
\hfill%
\noindent\parbox[t]{0.32\linewidth}{\centering\small\strut\textbf{(b)} Wave-action spectra for different $\mu$\par}%
\hfill%
\noindent\parbox[t]{0.32\linewidth}{\centering\small\strut\textbf{(c)} Wave-action spectra for different $L_c$\par}%
\par
\caption{Wave-action spectra of the wave field under different conditions.}
\label{fig:action_spectrum}
\end{figure}

At the initial time, the wave field consists of random noise, and the wave action is distributed approximately uniformly over a wide range of wavenumbers, so that the spectrum is rather flat. As the nonlinear evolution proceeds, modes with certain wavenumbers are selectively amplified, and the wave energy is gradually transferred from the broad background to a few discrete wavenumbers, which manifests itself as the appearance of peaks in the spectrum \cite{SotoCrespo2016}.

Figure~\ref{fig:action_spectrum}(a) shows the evolution of the wave-action spectrum $S_k$ with time. At the initial time ($t=0$), the spectrum is rather flat, and the wave action is distributed approximately uniformly over all wavenumbers, which corresponds to the characteristics of random noise. As the evolution proceeds to $t=25$ and $t=50$, the wave action concentrates in the background mode ($k=0$), and symmetric sidebands appear on both sides, which widen as the evolution proceeds.

Figure~\ref{fig:action_spectrum}(b) compares the wave-action spectra $S_k$ for different values of the standard deviation $\mu$, with the other conditions held fixed. It can be seen that, as $\mu$ increases, the wavenumber positions of the peaks remain essentially unchanged, whereas the peak heights increase significantly. This indicates that increasing $\mu$ does not change the characteristic wavenumbers that are selectively amplified, but rather enhances the amplification of the individual modes, so that the amplitude at the corresponding wavenumbers becomes larger. This result is also consistent with the earlier conclusion that the tail of the PDF rises with increasing $\mu$.

Figure~\ref{fig:action_spectrum}(c) compares the wave-action spectra $S_k$ for different correlation lengths $L_c$. As $L_c$ increases, the spectrum's width becomes narrower, and the wave action becomes more concentrated at small wavenumbers. This occurs because a larger $L_c$ implies a gentler spatial variation of the initial noise, which corresponds to a smaller characteristic wavenumber.

The wave action spectrum and the Lax spectrum characterize the wave field from different perspectives. The wave action spectrum describes the distribution of energy in the wavenumber domain, reflecting which frequency components are amplified. The Lax spectrum characterizes the type and number of nonlinear structures, reflecting what kinds of waves are formed. Together, the two provide a complete description of the evolution of integrable turbulence from frequency-domain and structural perspectives, respectively.

\subsection{Comparison with the NLS equation: role of the third-order dispersion and the derivative nonlinearity}
\label{sec:nls}

In order to examine the role of the derivative nonlinearity and of the third-order dispersion, the complex mKdV equation is compared in this subsection with the standard NLS equation,
\begin{equation}
iq_t + q_{xx} + 2|q|^2 q = 0,
\end{equation}
which is usually employed to describe situations in which the pulse lasts long and the waveform varies gently; however, when the pulse is as short as a few optical cycles, the third-order dispersion and the self-steepening effect can no longer be neglected \cite{Leblond2013}, and the complex mKdV equation becomes the appropriate model.

\begin{figure}[H]
\centering
\includegraphics[width=7.2cm,height=5.2cm,angle=0]{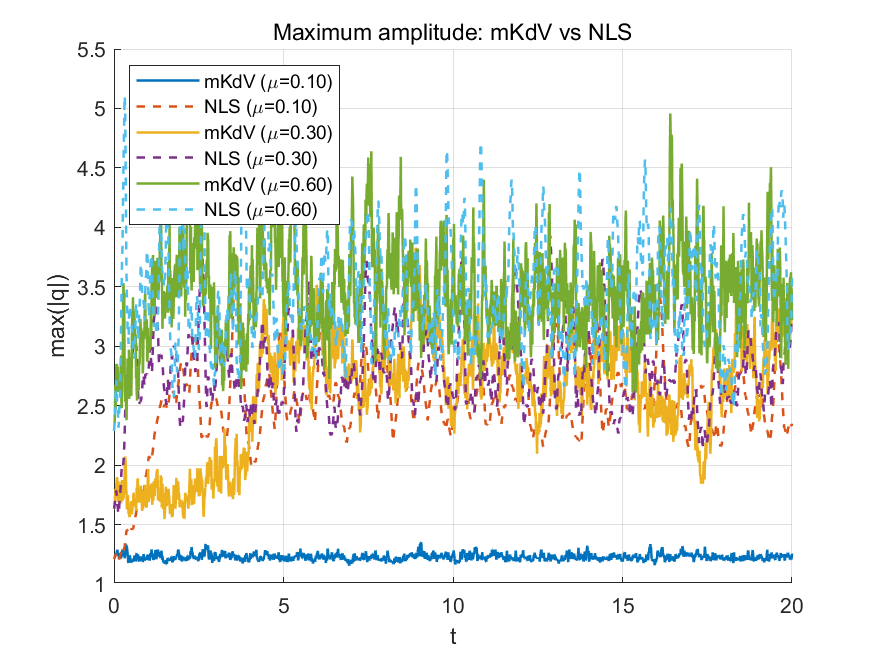}%
\hspace{0.3cm}%
\includegraphics[width=7.2cm,height=5.2cm,angle=0]{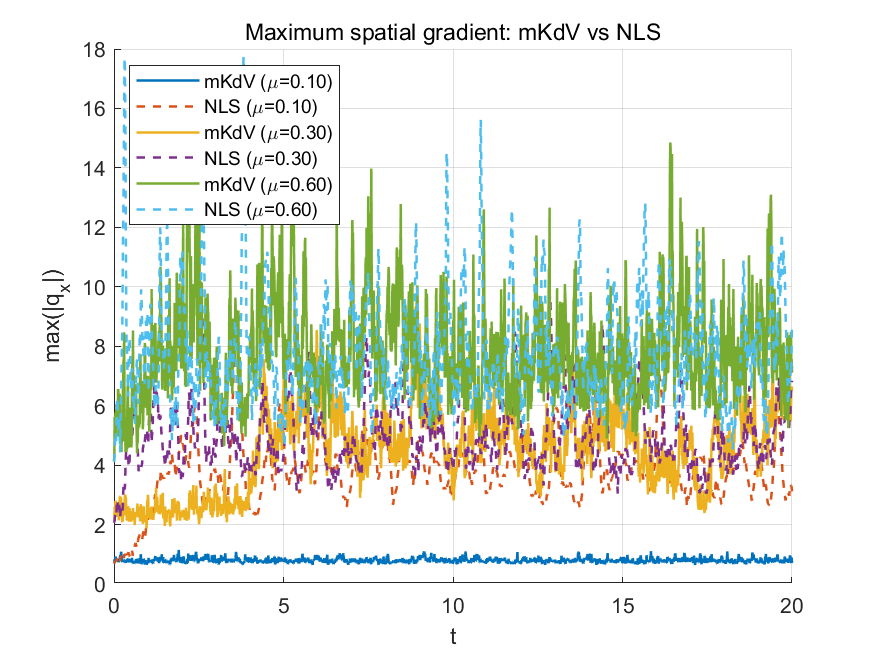}%
\par\smallskip
\parbox[t]{7.2cm}{\centering\small\strut\textbf{(a)} Evolution of the maximum amplitude $|q|_{\max}$\par}%
\hspace{0.3cm}%
\parbox[t]{7.2cm}{\centering\small\strut\textbf{(b)} Evolution of the maximum spatial gradient $|q_x|_{\max}$\par}%
\par
\caption{Comparison of the complex mKdV and standard NLS equations under identical initial conditions and with the same numerical scheme (solid lines: complex mKdV; dashed lines: NLS).}
\label{fig:nls_compare}
\end{figure}

To ensure a fair comparison, solve the two equations under identical initial conditions, $q(x,0)=c+\mu f(x)$, using the same numerical scheme (split-step Fourier and fourth-order Runge-Kutta) and the same resolution, so that any differences originate solely from the equations themselves. The results reveal fundamental differences between the two.

\textbf{(1) Different excitation mechanisms.} For $\mu=0.1$, the perturbation in the NLS equation is amplified exponentially within $t\approx2$ and reaches $|q|_{\max}\approx3$ because of the modulational instability of its constant background, whereas the maximum amplitude of the complex mKdV equation remains at about $1.25$ [Fig.~\ref{fig:nls_compare}(a)]. This is consistent with the conclusion of Section~\ref{sec:evolution}: the constant background of the complex mKdV equation is modulationally stable, and its integrable turbulence does not rely on a linear instability but is instead driven by the nonlinear evolution of a finite-amplitude random perturbation.

\textbf{(2) Comparable amplitudes under strong perturbations.} As $\mu$ increases to $0.6$, the maximum amplitudes attained by the two models become comparable, both reaching $3.5$--$4.9$.  It should be noted that at this point, the amplitude of the initial field itself is already close to $3$; therefore, this phenomenon stems in part from the contribution of the initial conditions under strong perturbations. However, the complex mKdV equation still grows from these initial conditions to $4.9$, indicating that nonlinear evolution does indeed play a role [Fig.~\ref{fig:nls_compare}(a)].

\textbf{(3) The self-steepening effect manifests itself in the spatial gradient rather than in the amplitude.} For $\mu=0.6$, the maximum spatial gradient of the complex mKdV equation is appreciably higher than that of the NLS equation (about $15$ versus $13$), whereas the maximum amplitudes of the two are comparable [Fig.~\ref{fig:nls_compare}(b)]. This indicates that the self-steepening effect arising from the third-order dispersion and derivative nonlinearity makes the waveform steeper and the spikes narrower, rather than simply increasing the amplitude of the wave field. Previous studies have shown that the self-steepening effect also significantly alters the statistical properties of extreme waves in integrable turbulence \cite{Zheng2025MNLS}, which corroborates the phenomenon observed here for the complex mKdV equation.

It is worth noting that previous studies have shown that, compared with the standard NLS equation, the chaotic wave field of the KE equation possesses a higher amplitude and a larger probability of rogue wave generation \cite{Li2024KE}. The present comparison shows, however, that this rule cannot be directly extended to the complex mKdV equation. As demonstrated in Section~\ref{sec:evolution}, the constant background of the complex mKdV equation is modulationally stable, and its integrable turbulence is driven by the nonlinear evolution of a finite-amplitude perturbation rather than by a linear instability. Consequently, under identical initial conditions, the amplitude of the complex mKdV equation grows more slowly than that of the NLS equation [Fig.~\ref{fig:nls_compare}(a)], and the role of its third-order dispersion and derivative nonlinearity is reflected in the steepness of the waveform rather than in its amplitude. Therefore, whether higher-order effects can enhance rogue wave generation depends crucially on whether the background is modulationally unstable.

In summary, compared with NLS-type equations, the third-order dispersion and derivative nonlinearity in the complex mKdV equation do not ``increase the probability of extreme events''; instead, they alter the excitation mechanism of integrable turbulence and appreciably increase the spatial steepness of the wave field. Therefore, the study of integrable turbulence in the complex mKdV equation cannot simply follow the conclusions obtained for NLS-type equations; it must instead start from the spectral structure and the nonlinear mechanism of the equation itself.

\section{Conclusions and discussion}
\label{sec:conclusion}

In this paper, we have studied integrable turbulence and the statistical properties of the chaotic wave field of the complex mKdV equation. We first
computed the eigenvalue spectra of the soliton, breather and rogue-wave solutions numerically by the Fourier collocation method, in order to establish
the correspondence between the spectral distribution and the different nonlinear structures. The results show that a soliton corresponds to a pair of
complex-conjugate eigenvalues with a nonzero real part, a breather to a pair of complex-conjugate eigenvalues on the imaginary axis, and a rogue wave to the
critical points $\lambda=\pm ic$ on the imaginary axis.

We then simulated the complex mKdV equation starting from a plane wave with superimposed random noise and generated a chaotic wave field. The evolution of
the initial field intensity and of the maximum amplitude with time shows that a larger standard deviation $\mu$ leads to a stronger chaotic deviation, a faster
approach to the chaotic state and a higher peak amplitude. The PDFs at a late stage of the evolution show that their tails rise markedly with increasing
$\mu$, indicating a significant increase in the probability of rogue-wave generation. The changes in the eigenvalue and wave-action spectra identify a
transition from breather turbulence to soliton turbulence and confirm the appearance of large-amplitude spike structures. As for the correlation length
$L_c$, an increase in $L_c$ enlarges the spatial scale of the structures and raises the PDF tail; the eigenvalue statistics further show that the fraction
of solitons decreases while that of breathers increases, but solitons remain dominant throughout, so that the turbulence type is not changed. The
correlation length therefore mainly affects the spatial scale of the wave field and adjusts the relative proportions of the two types of structure.

To reveal the role of the third-order dispersion and of the derivative nonlinearity, we compared the complex mKdV equation with the standard NLS
equation under identical initial conditions. Since the constant background of the complex mKdV equation is modulationally stable, its integrable turbulence
does not rely on MI. For weak perturbations its amplitude grows far more slowly than that of the NLS equation, whereas for strong perturbations the two
amplitudes become comparable while the complex mKdV equation exhibits an appreciably higher maximum spatial gradient. The role of the derivative
nonlinearity and of the third-order dispersion is therefore not to raise the probability of extreme events but to alter the excitation mechanism of
integrable turbulence and to make the waveform steeper and the spikes narrower. These results contribute to a numerical and statistical understanding of
integrable turbulence and of rogue-wave formation in the complex mKdV equation, and they provide a useful reference for studies of waveform steepening and
spectral broadening in ultrashort-pulse propagation.

\section*{Acknowledgements}

 W. Q. Peng is supported by the  National Natural Science Foundation of China (No. 12501335), Natural Science
Foundation of Shandong Province (No. ZR2025QC1464, No. ZR2026MS0081), Fundamental Research Funds for the Central Universities
(No. 202513017) and Qingdao Postdoctoral Science Foundation (No. QDBSH20250102147).

%

\setcounter{equation}{0}
\renewcommand\theequation{A.\arabic{equation}}

\bibliographystyle{unsrt}
\bibliography{references}

\end{document}